\pdfoutput=1
\RequirePackage{fix-cm}
\documentclass[smallextended]{svjour3}       
\smartqed  
\usepackage{graphicx}
\usepackage{natbib}
\usepackage{booktabs} 
\usepackage{amsmath,amssymb}
\usepackage{graphicx}
\usepackage{url}
\usepackage[ruled,vlined]{algorithm2e}
\usepackage{multirow,cite}
\usepackage{mwe}
\usepackage{subfig}
\usepackage{color}
\DeclareMathOperator{\prox}{prox}
\begin{document}

\title{TrendNets: Mapping Emerging Research Trends From Dynamic Co-Word Networks via Sparse Representation}

\titlerunning{TrendNets}        

\author{Marie Katsurai \and Shunsuke Ono}


\institute{M. Katsurai \at
              Doshisha University\\
              1-3 Tatara Miyakodani, Kyotanabe-shi, Kyoto \\
              Tel.: +81-774-65-7687\\
              \email{katsurai@mm.doshisha.ac.jp}\\
              ORCID: 0000-0003-4899-2427
\and
  S. Ono \at Tokyo Institute of Technology\\
  4259 Nagatsuta-cho, Midori-ku, Yokohama, Kanagawa\\
  \email{ono@c.titech.ac.jp}
}

\date{Received: date / Accepted: date}

\maketitle

\begin{abstract}
Mapping the knowledge structure from word co-occurrences in a collection of academic papers has been widely used to provide insight into the topic evolution in an arbitrary research field.
In a traditional approach, the paper collection is first divided into temporal subsets, and then a co-word network is independently depicted in a 2D map to characterize each period's trend.
To effectively map emerging research trends from such a time-series of co-word networks,
this paper presents TrendNets, a novel visualization methodology that highlights the rapid changes in edge weights over time.
Specifically, we formulated a new convex optimization framework that decomposes the matrix constructed from dynamic co-word networks
into a smooth part and a sparse part:
the former represents stationary research topics, while the latter corresponds to bursty research topics.
Simulation results on synthetic data demonstrated that our matrix decomposition approach achieved the best burst detection performance over four baseline methods.
In experiments conducted using papers published in the past 16 years at three conferences in different fields,
we showed the effectiveness of TrendNets compared to the traditional co-word representation.
We have made our codes available on the Web to encourage scientific mapping in all research fields.
\keywords{emerging research trends \and dynamic co-word networks \and burst detection \and science mapping}
\end{abstract}

\section*{Article Highlights}
\begin{itemize}
 \item We present a novel emerging research trend mapping method that can be easily introduced to traditional co-word analysis. 
 \item The proposed matrix decomposition approach extracts sparse networks showing bursty research topics, which are named TrendNets.
 \item Experiments conducted using synthetic and real-world datasets demonstrate the effectiveness of TrendNets.
\end{itemize}

\section*{Introduction}

Due to the continuous increase of journals and conferences,
as well as the wide spread of preprint servers such as arXiv\footnote{https://arxiv.org/},
a massive amount of academic papers has accumulated.
Revealing new research topics can bring significant benefits for people involved in the research environment, including researchers, research administrators, publishers, and funding bodies.
However, it is hard even for experts to keep up with the large number of papers published.

To facilitate an understanding of an arbitrary research field,
visualization of emerging research topics using papers published in the target field
has attracted much attention~\citep{83-callon-ssi,12-assefa-jasist,12-munoz-quality,12-ronda-smj,12-zhang-plos,13-hu-sci,10-li-tits,16-topalli-jwb}.
One widely used technique is \textit{co-word analysis}~\citep{83-callon-ssi}, which maps the knowledge structure from word co-occurrences in the papers.
In a traditional approach, the paper collection is first divided into temporal subsets (e.g., yearly or multiple years) according to the publication dates of the papers.
Then, focusing on words in the content of papers such as titles and abstracts, the co-occurrence frequency of each word pair is calculated by counting the number of papers in which the two words appear together.
Finally, the knowledge structure of each time period is visualized as a co-word network whose nodes correspoind to words and whose edges reflect high co-occurrence frequencies between words, and in which the number of edges in the network is generally determined by thresholding edge weights.
A series of co-word networks over time (called dynamic co-word networks) helps us to investigate how a target field has developed~\citep{12-assefa-jasist,12-munoz-quality,12-ronda-smj,12-zhang-plos,13-hu-sci,10-li-tits,16-topalli-jwb}.
However, the traditional approach does not have the means to highlight how the emerging research trends change from time period to period;
it might depict edges, even if the corresponding word associations are stable or gradually become stronger over time.
Such stable word pairs can disturb the finding of discriminative or emerging research topics that characterize the time period.
Although there are several methods with which to visualize temporal changes of word clusters or topics~\citep{14-wang-sci,14-song-sci,05-mei-kdd,08-hall-emnlp}, these usually describe dominant research topics in a time period using general and frequent single words, rather than specific technical terms that suddenly attract attention.

To make it easier to understand the research topic evolution of a target field,
this paper presents TrendNets, a novel emerging research trend visualization approach that detects rapid changes in edge weights in dynamic co-word networks.
Our aim here is to find ``bursty research topics''---topics that have been growing rapidly rather than topics that are merely popular. That is, a research topic is considered to be bursty in a given time period if it is heavily discussed in that time period, but not in the previous time periods. To measure the burstiness of a research topic, we calculate the difference in word co-occurrence frequency between two successive time periods.
In the proposed approach, we first arrange the vectorized edge weights of co-word networks in columns of a single matrix in the order of time.
Then, to detect bursty research topics, we formulate a convex optimization framework that decomposes the matrix 
into a smooth part corresponding to stationary topics and a sparse part corresponding to bursty topics in a computationally efficient manner.
We term this framework \textit{Fast Sparse-Smooth Matrix Decomposition}. 
After this optimization, the non-zero entries of the resulting sparse networks are considered to represent bursty topics for the corresponding time periods. 
Experiments on both toy data and real publication data show that TrendNets can effectively detect emerging research trends.

The main contributions of this paper can be summarized as follows:
\begin{itemize}
 \item We introduce a novel bursty research topic detection scheme for a traditional co-word analysis for visualizing emerging research trends.
       The proposed method is computationally efficient, even for a large matrix calculated from dynamic co-word networks.

 \item We perform extensive experiments based on both synthetic and real-world datasets, which 
show that TrendNets effectively describes the bursty topics when compared with the conventional co-word representations. 

\item Our method can automatically detect technical phrases from a pool of title words without predefined vocabulary, which is useful because obtaining a set of titles is much easier than collecting keywords to papers provided by authors to papers.

 \item We have made our codes available to encourage science mapping in all disciplines.
\end{itemize}

The remainder of this paper is organized as follows:
the next section provides a brief review of the related studies;
the third section presents the details of TrendNets;
the results of the experiments on the synthetic dataset and paper collections are presented in the fourth and fifth sections, respectively;
finally, the paper is summarized and some possible directions for future work are suggested.
\section*{Related Work}

As scholarly big data is growing, science mapping has become more important for encouraging the activities of researchers, research administrators, and science policymakers.
Many studies construct knowledge networks using the bibliographic metadata of papers
to visualize the relationships between academic-related entities such as authors, papers, and technical terms.
For example, co-authorship networks have been exploited to find influential researchers and research groups~\citep{05-borner-comp}.
Constructing citation or co-citation networks is also a popular approach for visualizing content-based relevance among research subfields~\citep{08-shibata-tech}.
However, it is difficult for citation-based methods to grasp the latest trend because papers usually require time to be cited by other papers.
On the other hand, co-word networks have the advantage of being able to discover emerging technologies in a timely manner because they can be constructed as quickly as new papers are published.
Using textual words can directly map the knowledge structure of a research field into a conceptual space.
The visualization of word associations in a network form is known to provide intuitional understanding~\citep{02-motter-phi,13-drieger-pro,99-doerfel-hcr}.
Thus, co-word analysis has been widely used to reveal research advances in several fields, including science education~\citep{12-assefa-jasist}, strategic management~\citep{12-ronda-smj}, patience adherence~\citep{12-zhang-plos}, and information retrieval~\citep{13-hu-sci}.
It has also been used to characterize the publication trends of journals~\citep{15-ravikumar-sci} or conferences~\citep{14-liu-chi}.
In conventional visualization,
the number of edges in a network is generally determined via the thresholding of edge weights (i.e., word co-occurrence frequency).
However, this simple approach tends to produce a lot of word pairs that are stable over time.

To visualize the transition of research topics in dynamic co-word networks, there are methods that detect clusters in each co-word network separately and then track the clusters over time~\citep{14-wang-sci,14-song-sci}.
Specifically, \citet{14-wang-sci} applied community detection to a co-word network at each time period and connected communities based on their node-based or edge-based similarity across successive time periods.
Song et al.~\citep{14-song-sci} clustered words via the Markov Random Field and tracked each cluster using word similarity to show its development.
These conventional methods can visualize the evolution of representative semantic clusters in a target research field.
Because the clusters are usually labeled with general words,
it is difficult to highlight specific technical terms that are emerging and likely to form a new technology.

Another line of topic evolution identification has leveraged probabilistic topic models
such as Latent Dirichlet Allocation (LDA)~\citep{03-blei-jmlr}.
For example, \citet{05-mei-kdd}  applied a topic model to papers published in each time period separately and then linked the obtained topics across successive periods based on their similarities.
\citet{08-hall-emnlp} applied LDA to an entire collection of papers to model research topics and then plotted the number of papers assigned to each topic within each time period.
There are also advanced models that can directly treat papers' time stamps in topic modeling~\citep{06-wang-kdd,06-blei-icml,08-wang-uai}.
These topic model-based methods usually label resulting clusters with top frequent words within the clusters, which does not emphasize emergent technical terms.
On the other hand, TrendNets considers how the co-occurrence frequency of a word pair suddenly increases from previous time periods,
directly extracting temporally representative word pairs.


Several studies have regarded words displaying a sudden increase in usage to be indicative of emerging research trends~\citep{04-mane-pnas,15-kim-sci}. 
Traditional burst detection approaches include the thresholding of occurrences (e.g., the moving average~\citep{04-vlachos-sigmod})
and state transition modeling (e.g., a two-state automaton proposed by \citet{03-kleinberg-kdd}).
One of the most famous science mapping tools is named CiteSpace~\citep{06-chen-jasist},
and it implements the Kleinberg's burst detection method to show representative words in a specific time period.
Kleinberg's method requires two parameters that are difficult to simultaneously tune.
To increase the usefulness of science mapping, our method for constructing TrendNets is formulated to work with only a single parameter.
In addition, the related works usually assume that research keywords of papers are obtainable for input data. However, author-provided keywords of conference papers are not always available as open bibliographic information. Thus, we opted to use paper titles only, which are much easier to collect than keywords. Because our method focuses on the burst of edge weights in co-word networks rather than single-word occurrences, we can automatically extract emergent research phrases/terms that are composed of multiple words in the network form. In our experiments, we demonstrate that our TrendNets can effectively organize meaningful phrases from single words within paper titles.

\section*{Proposed Method}

\begin{figure*}[t]
 \centering
 \includegraphics[clip,width=1.0\textwidth]{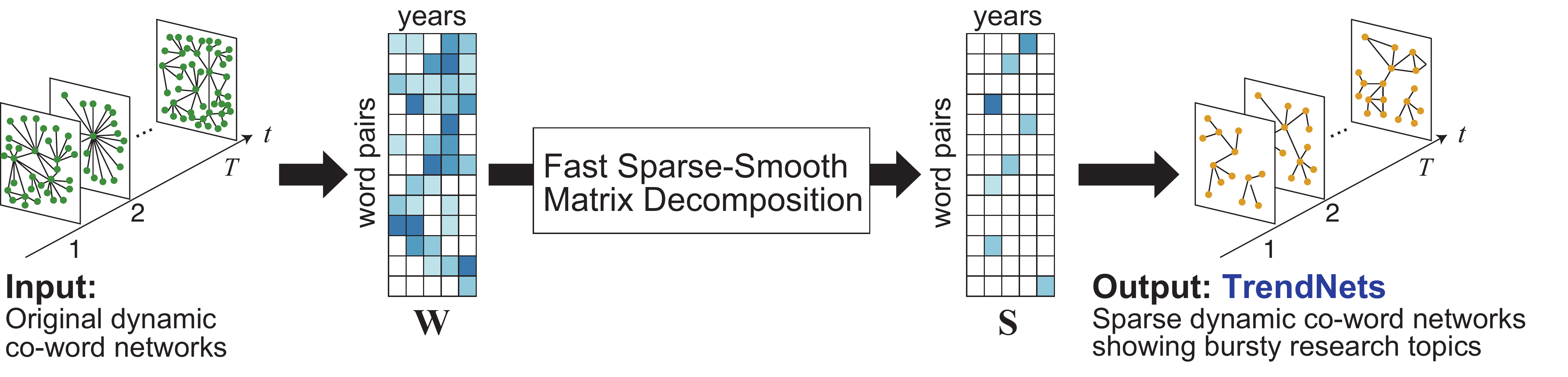}
 \caption{
 An overview of the proposed method that constructs TrendNets.
 \label{fig:overview}
 }
\end{figure*}

This section describes how to construct TrendNets to map emerging research trends from dynamic co-word networks.
An overview of the proposed method is shown in Fig.~\ref{fig:overview}.
As shown, we first construct a co-word network for each time period and convert its edge weights into a single matrix whose rows and columns correspond to word pair indexes and time period indexes, respectively. Then, we decompose the matrix it into a smooth part and a burst part. According to our definition of bursty research topics (see Introduction), the matrix decomposition step calculates the difference in word co-occurrence frequency between two successive time periods. The burst part corresponds to a large increase for a word pair, which is represented as a sparse matrix. Finally, the resulting sparse matrix is rearranged into a time-series of sparse co-word networks, in which each of the networks visualize bursty topics during the corresponding period. The details of each procedure are described below.

\subsection*{Co-word network construction}

\begin{figure}[t]
 \centering
 \includegraphics[clip,width=0.8\textwidth]{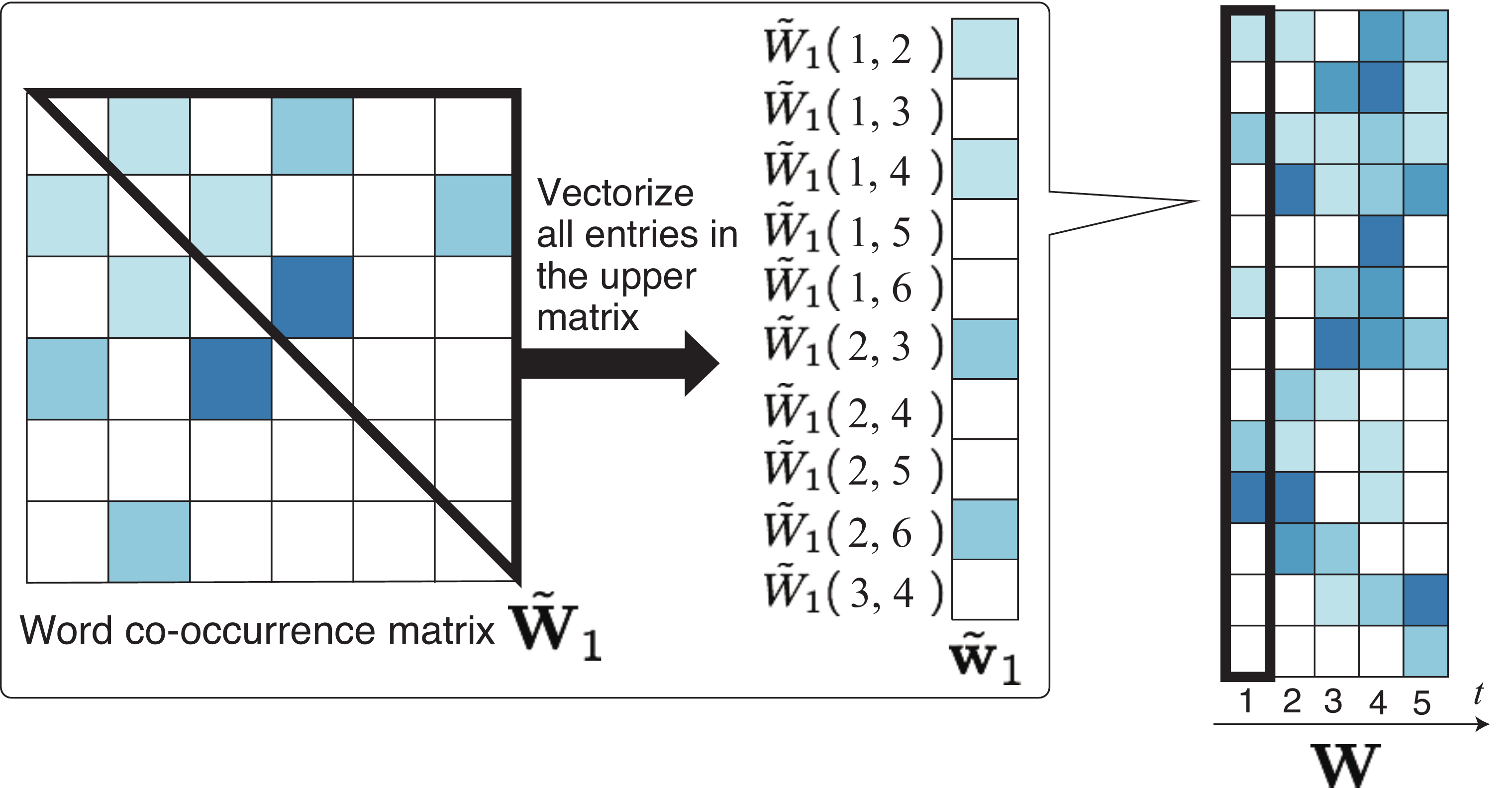}
 \caption{
 Matrix construction from a time series of co-word graphs.
 \label{fig:matrix}
 }
\end{figure}

Let $\Omega$ be a collection of papers published in an arbitrary research field. 
We denote the number of unique words in $\Omega$ by $M$.
According to the publication dates of the papers,
we divide the paper collection $\Omega$ into exclusive temporal subsets $\Omega(1), \Omega(2), \cdots, \Omega(T)$
such that $\Omega = \cup_{t=1}^T \Omega(t)$.
To construct a co-word network at the $t$-th time period ($t\in~\{1,2,\cdots, T\}$),
following \citet{14-liu-chi},
we calculate how often two words appear on the same paper on average during the period as follows:
\begin{align}
 \tilde{W}_t(i, j) = \frac{n_t(i, j)}{|\Omega(t)|}, i\neq j, i, j\in \{1, 2, \cdots, M\}, \label{eq:weight}
\end{align}
where $n_t(i, j)$ is the number of papers in which the $i$-th and $j$-th words appear together during the $t$-th time period
and $|\cdot|$ represents the cardinality of the set.
This forms a symmetrical matrix $\tilde{\mathbf{W}}_t\in \mathbb{R}^{M\times M}$, which corresponds to an adjacency matrix of a conventional co-word network.
Then, as shown in Fig.~\ref{fig:matrix},
we turn the upper triangular part of $\tilde{\mathbf{W}}_t$ into a column vector $\tilde{\mathbf{w}}_t\in \mathbb{R}^{N}$,
in which $N = M(M-1)/2$.
Finally, we arrange all vectors over time in the columns
to construct a single matrix as $\mathbf{W} = [\mathbf{w}_1, \mathbf{w}_2, \cdots, \mathbf{w}_T] \in \mathbb{R}^{N\times T}$.

\subsection*{Fast Sparse-Smooth Matrix Decomposition}

In what follows, we extract sparse graphs from the time series of co-word networks
by solving a newly formulated convex optimization problem. 
Specifically, we decompose a given $\mathbf{W}$ into the sum of a sparse matrix and a smooth matrix, 
in which the sparse matrix is expected to only contain rapidly changed entries that are closely related to bursty topics. The convex optimization problem is given as follows:
\begin{equation}\label{prob:sparse}
\min_{\mathbf{S}}  \frac{1}{2}\|D(\mathbf{W} - \mathbf{S})\|_F^2 + \lambda\|\mathbf{S}\|_1, 
\end{equation}
where $\mathbf{S}$ contains extracted sparse graphs in its columns,
$D$ is a linear operator computing column differences,
and $\|\cdot\|_1$ and $\|\cdot\|_F$ are the $\ell_1$ and Frobenius norms, respectively. 
The operator $D$ corresponds to the calculation of the difference in word co-occurrence frequency between two successive time periods.
In Prob.~\eqref{prob:sparse}, the first term plays a role in promoting the sparsity of $\mathbf{S}$
(NOTE: the $\ell_1$ norm is the tightest convex relaxation of the $\ell_0$ pseudo-norm, i.e., a reasonable sparsity measure).
Meanwhile, the second term keeps the remaining matrix $\mathbf{W}-\mathbf{S}$ smooth in the column direction,
implying that the matrix is expected to consist of entries that are not greatly changed during any periods.

Because Prob.~\eqref{prob:sparse} is a convex but nonsmooth optimization problem
and, indeed, has no closed-form solution, we have to use some iterative algorithms to solve it.
In our framework,
the fast iterative shrinkage-thresholding algorithm (FISTA) \citep{09-beck-siam} is adopted.
FISTA can solve convex optimization problems in the following form:
\begin{equation}\label{prob:FISTA}
\min_{\mathbf{x}} f(\mathbf{x}) + g(\mathbf{x}),
\end{equation}
where $f$ is a smooth convex function with a Lipschitzian gradient,
and $g$ is a possibly nonsmooth convex function, and its proximity operator\footnote{The proximity operator of index $\gamma>0$ of a proper lower semicontinuous convex function $h$
is defined by $\prox_{\gamma h}(\mathbf{x}):=\arg\min_{\mathbf{y}}h(\mathbf{y})+\tfrac{1}{2\gamma}\|\mathbf{y}-\mathbf{x}\|^2$. }
\citep{61-moreau-cr} is available.
The algorithm is given as follows:
for an initial vector $\mathbf{y}_0$ and $z_0=1$, iterate
\begin{align}
\left\lfloor\begin{array}{l}
\mathbf{x}_{k+1}:=\prox_{\frac{1}{L} g}(\mathbf{y}_{k}-\frac{1}{L}\nabla f(\mathbf{y}_{k})),\\
z_{k+1}:= \frac{1 + \sqrt{1 + 4z_{k}^2}}{2},\\
\mathbf{y}_{k+1}:= \mathbf{x}_{k} + \frac{z_{k}-1}{z_{k+1}}(\mathbf{x}_{k}-\mathbf{x}_{k-1}),
\end{array}\right.\label{alg:FISTA}
\end{align}
where $k$ is the iteration number and
$L$ is the Lipschitz constant of $\nabla f$.

Let $f:=\frac{1}{2}\|D(\cdot - \mathbf{S})\|_F^2$
and $g:=\lambda\|\cdot\|_1$.
Then, $f$ is clearly differentiable, and 
its gradient:
\begin{equation*}
\nabla f(\mathbf{S}) = -\mathbf{D}^\top\mathbf{D}(\mathbf{W}-\mathbf{S})
\end{equation*}
has the Lipschitz constant $L$ being equal to $\|D\|_{op}^2$,
where~$\|\cdot\|_{op}$ is the operator norm.
Meanwhile, since $g$ is the $\ell_1$ norm, its proximity operator boils down to the soft-thresholding operation:
\begin{equation}\label{softthresh}
[\prox{\tfrac{1}{L} g}(\mathbf{S})]_{i,j} = \mbox{sgn}(S_{i,j})\max\{S_{i,j}-\tfrac{\lambda}{L}, 0\}.
\end{equation}
As a result, FISTA can be readily applied to Prob.~\eqref{prob:sparse},
which is summarized in Alg.~\ref{alg:fSSMD}.
\begin{algorithm}[h]
\LinesNumbered
\SetKwInOut{Input}{input}
\SetKwInOut{Output}{output}
\caption{Fast sparse-smooth matrix decomposition by FISTA}
\label{alg:fSSMD}
\Input{$\mathbf{Y}_0=\mathbf{W}$, $z_0=1$, $L=\|D\|_{op}^2$, $\lambda$, $k=0$}
\While{A stopping criterion is not satisfied}{
$\mathbf{S}_{k+1}=\mathbf{Y}_{k}+\frac{1}{L}D^{*}D(\mathbf{W}-\mathbf{Y}_k)$\;
$\mathbf{S}_{k+1}\leftarrow\mbox{soft-thresholding}(\mathbf{S}_k,\frac{\lambda}{L})$ by \eqref{softthresh}\;
$z_{k+1}:= \frac{1 + \sqrt{1 + 4z_{k}^2}}{2}$\;
$\mathbf{Y}_{k+1}:= \mathbf{S}_{k} + \frac{z_{k}-1}{z_{k+1}}(\mathbf{S}_{k}-\mathbf{S}_{k-1})$\;
$k\leftarrow k+1$\;
}
\Output{$\mathbf{S}_k$}
\label{algo_disjdecomp}
\end{algorithm}

FISTA is a reasonable choice for solving Prob.~\eqref{prob:sparse},
and the reasons are twofold:
(i) its convergence rate is $\mathcal{O}(1/k^2)$ \citep{09-beck-siam}, i.e., very fast,
and (ii) it uses only the gradient of $f$ and the proximity operator of $g$,
and, in our case, these can be computed efficiently
as explained above (the computational cost is linear with the number of entries in $\mathbf{W}$).
These properties are important in our framework because matrix $\mathbf{W}$ is large,
and we must avoid expensive procedures, such as singular value decomposition.
In all experiments, we set the Lipschitz constant $L$ in Eq.~\eqref{alg:FISTA} to 4.0\footnote{We conducted a preliminary experiment and observed that our matrix decomposition is not sensitive to $L$.}.

\subsection*{Visualization of bursty research topics}

After solving \eqref{prob:sparse},
we transform the $t$-th column in the resulting sparse matrix $\mathbf{S}$ into a symmetrical matrix $\tilde{\mathbf{S}}_t\in \mathbb{R}^{M\times M}$.
The non-zero, positive elements of $\tilde{\mathbf{S}}_t$ form a new co-word network for the $t$-th time period
in which nodes represent words and edges associate the words to form bursty research topics.
For the effective visualization of the research topics, 
nodes in the graph are generally grouped according to the graph structure~\citep{14-liu-chi,16-silva-ji}, in which spectral clustering or community detection techniques are often exploited.
For this paper, we use the Louvain method~\citep{08-blondel-jsm} to find semantic clusters of words.
\section*{Experiment I: Simulation on Synthetic Data}

\subsection*{Synthetic dataset construction}

We first quantitatively evaluated the performance of our matrix decomposition method that enables TrendNets.
Because a ground truth set for bursty research topic detection is unavailable, 
we constructed a synthetic dataset using the following two steps:

\begin{itemize}
 \item{\textbf{Constructing stable co-word networks}.} 
We first generated 50 ``stable'' co-word networks,
which are mostly similar to each other on the basis of a statistical model.
Specifically, we used a subset of documents in the Reuters-21578 corpus\footnote{\url{http://www.daviddlewis.com/resources/testcollections/reuters21578/}}
that were tagged with ``trade''
to train a 5-gram language model.
For each time period $t \in \{1, 2, \cdots, 50\}$,
the 5-gram language model produced a set of 5,000 documents that were statistically identical to the given corpus
in which the maximum number of words in each document was set to 12.
The document set constructed for $t$ is denoted by $\Omega(t)$.
Discarding words that appear less than 30 times in $\{\Omega(t)\}_{t=1}^{50}$
resulted in a vocabulary of $M=6,696$ unique words.
We counted the co-occurrence frequency of the $i$-th and $j$-th words in $\Omega(t)$, which is denoted by $G_t(i, j)$,
producing a co-word network with adjacency matrix $G_t\in \mathbb{R}^{M\times M}$.
The resulting time series of networks can be considered as being stable over time.
For each word pair $(i, j)$, the mean and standard deviation values of $\{G_t(i, j)\}_{t=1}^{50}$ are denoted by $\mu(i, j)$ and $\sigma(i, j)$, respectively.

\begin{figure}[t]
 \centering
 \begin{tabular}{cc}
  \includegraphics[clip,width=0.4\textwidth]{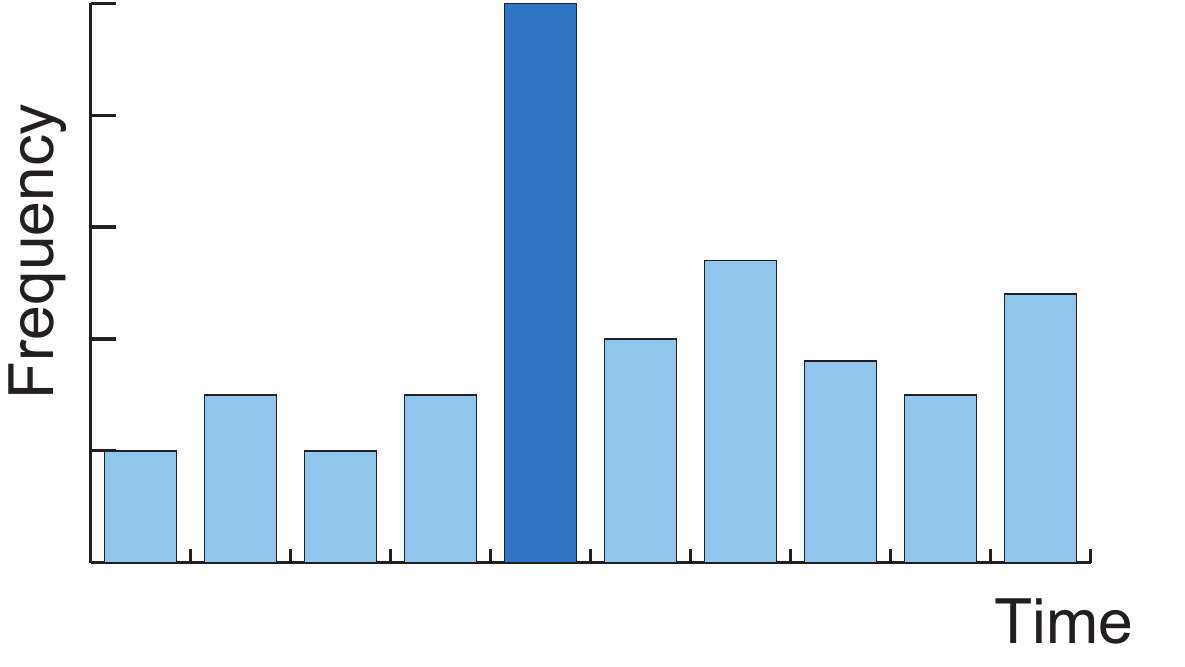}
  & 
  \includegraphics[clip,width=0.4\textwidth]{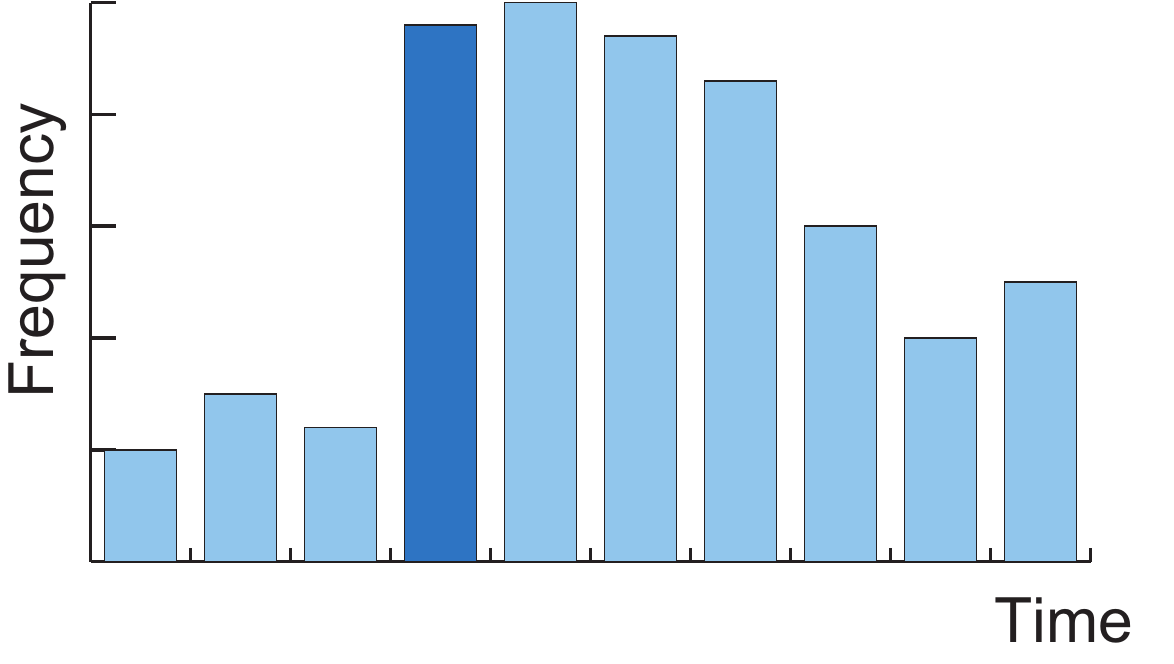}
  \\
 (a) & (b)
 \end{tabular}
 \caption{
 Illustrations of two types of bursts generated for the synthetic dataset. 
 (a) Type-A bursts: word pairs that suddenly increase and disappear later;
 (b) Type-B bursts: word pairs that suddenly increase and keep the co-occurrence frequency after that.
 Deep blue bars indicate time periods that should be detected as bursts.
 \label{fig:Toydata}
 }
\end{figure}

\item{\textbf{Generating synthetic bursts}.}
Next, we artificially generated bursts on the time series of networks to construct ground truth data.
As shown in Fig.~\ref{fig:Toydata}, we considered the following two types of bursts:
``Type-A bursts'' (Fig.~\ref{fig:Toydata}a) are word pairs that have a peak of frequency at a certain time
and ``Type-B bursts'' (Fig.~\ref{fig:Toydata}b) are word pairs that keep the high frequency after a sudden increase.
To generate these bursts, for each time period $t$,
we randomly picked a word pair $(i', j')$ with probability $p=0.05$ from all pairs such that $\mu(i', j')>3$
and regarded the tuple $(i', j', t)$ as a burst point.
We further randomly classified each word pair into Type-A or Type-B with probability $p=0.95$ and $p=0.05$, respectively.

For a burst point $(i', j', t)$ labeled with Type-A, we increased its edge weight as follows:
\begin{align}
 G_t(i', j') \leftarrow \mu(i', j') + u_t(i', j')\sigma(i', j'),\label{eq:burst}
\end{align}
where $u_t(i', j')$ is a random number drawn uniformly from $[3, 6]$.

For a burst point $(i', j', t)$ labeled with Type-B, we iteratively computed $G_{t'}(i', j')$ using Eq.~\eqref{eq:burst} for $t'=t,t+1,\cdots, 50$.

Finally, we ran the proposed method starting from Eq.~\eqref{eq:weight}, in which $n_t(i, j)=G_t(i,j)$ and $|\Omega(t)|=50$, respectively.

\end{itemize}

\subsection*{Baselines}

In the experiment, the proposed method was compared with the following four baseline methods:
\begin{itemize}
 \item \textbf{Baseline 1: Thresholding}.
 This approach identifies word pairs whose co-occurrence frequencies exceed a predetermined threshold value as bursts.
 That is,  if $\tilde{W}_t(i, j)$ in Eq.~\eqref{eq:weight} is larger than threshold $\tau_1$,
 then the word pair $(i, j)$ is detected as a burst at $t$.
 Baseline 1 corresponds to the traditional co-word network visualization.

 \item \textbf{Baseline 2: Thresholding of time derivative}.
 This approach regards word pairs whose time derivatives of frequency exceed a predetermined threshold value as bursts.
 That is, if $\tilde{W}_t(i, j) - \tilde{W}_{t-1}(i, j)$ is larger than threshold $\tau_2$,
 then the word pair is marked as a burst at $t$.

 \item \textbf{Baseline 3: Thresholding of differences from the average over time}.
 This approach detects word pairs if the difference between those frequencies and the average frequency over time exceeds a predetermined threshold value.
 That is, if $\tilde{W}_t(i, j) - \mu(i,j)$ is larger than threshold $\tau_3$,
 then the word pair is marked as a burst at $t$.

 \item \textbf{Baseline 4: Kleinberg's method}.
 This approach exploits a well-known burst detection model presented by Kleinberg~\citep{03-kleinberg-kdd}
and is implemented to visualize emerging research trends in conventional studies~\citep{06-chen-jasist,17-katsurai-icbda}.
 It models a time series of a word pair's co-occurrence frequency as a finite-state automaton that consists of a base state and a burst state.
 The burst degree of the word pair at each time period was calculated from the state transition sequence.
 The parameters required in this model are $s$ and $\gamma$.
 We used $s=2.0$ and changed the value of $\gamma$.
\end{itemize}

\begin{figure}[t]
 \centering
 \includegraphics[clip,width=0.8\textwidth]{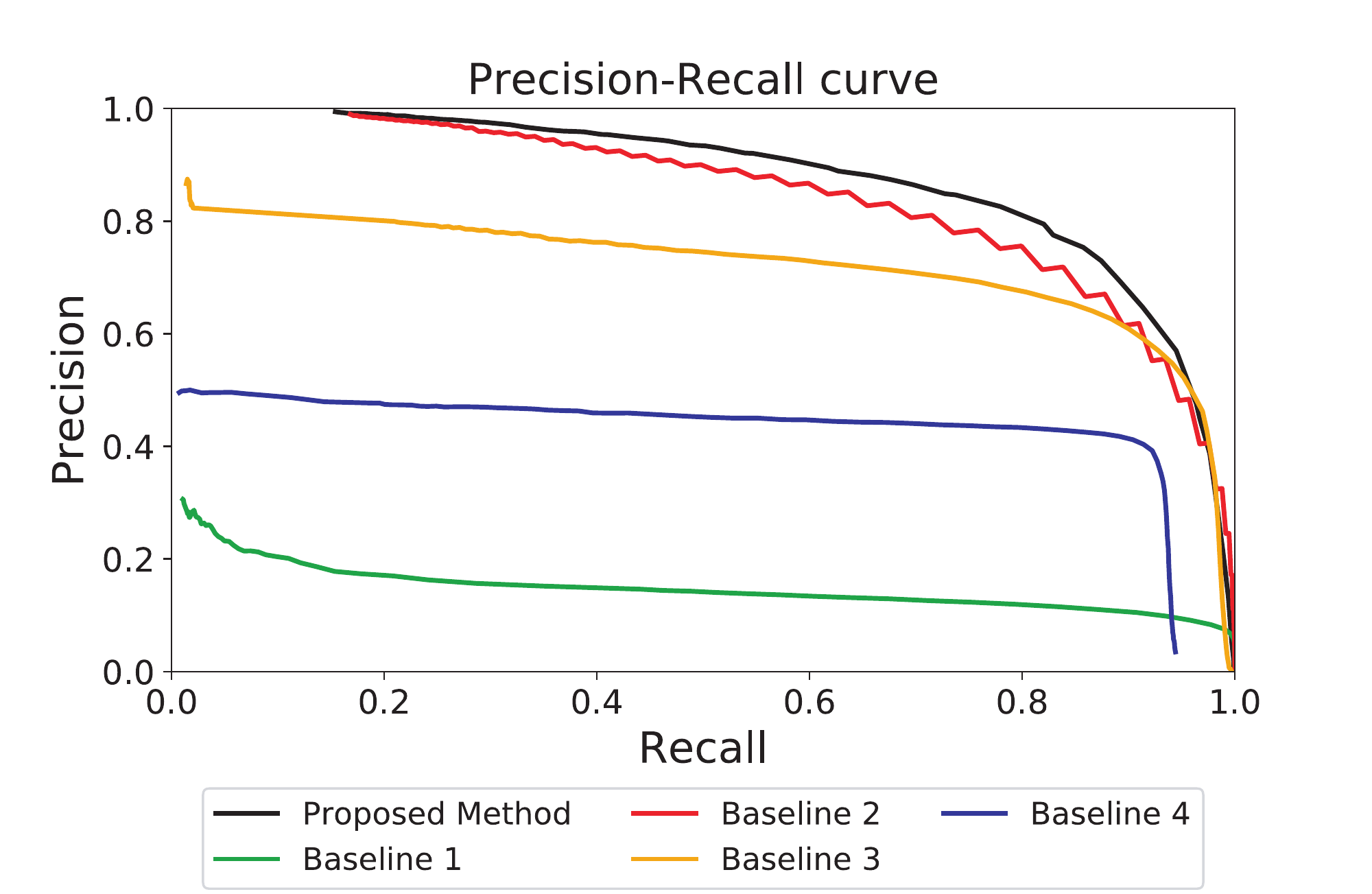}
 \caption{
 PR curve for burst detection results on synthetic data. \label{fig:PR_curve}
 }
\end{figure}

\begin{table}[t]
 \caption{AUC measured using burst detection results on synthetic data.
 \label{table:AUC}
 }
 \centering
 \begin{tabular}[t]{ccccc}\hline
  {Proposed Method} & {Baseline 1} & {Baseline 2} & {Baseline 3}& {Baseline 4} \\ \hline
  \textbf{0.876} & {0.148} & {0.845} & {0.720}  & {0.427}\\
 \hline
 \end{tabular}
\end{table}

\subsection*{Simulation results}
After applying each method to synthetic data,
we calculated the precision and recall measures as follows:
\begin{align}
 Recall &= \frac{\text{\# of correctly detected bursts}}{\text{\# of synthetic bursts}},\\
 Precision &= \frac{\text{\# of correctly detected bursts}}{\text{\# of detected bursts}}.
\end{align}
Figure~\ref{fig:PR_curve} shows the Precision-Recall (PR) curve for each method,
while Table~\ref{table:AUC} shows the Area Under the PR Curve (AUC).
As shown, we found that the proposed method achieved the best burst detection performance.
Baseline 1 provided the worst performance,
which implies that the traditional co-word visualization might not highlight an emerging trend in a certain time period.
Baseline 4, an application of Kleinberg's burst detection, achieved a relatively better performance; however, there is unfortunately no automatic method for determining its two parameters $s$ and $\gamma$.
Our burst detection method is more practical because it works with only a single parameter $\lambda$, which changes the sparsity of networks.
Baseline 2, that calculates the time derivative,
is the most similar to our method because both aim to detect a sudden increase in edge weights of dynamic co-word networks.
However, Baseline 2 often gives false detection of word pairs associated with large values of $\mu(i, j)$
because the time derivative of such word pairs tends to become relatively large.
Our method solved this problem by considering the smoothness along the time direction for each word pair.

\section*{Experiment II: TrendNets Construction Using Conference Papers}

This section presents the results of the experiments on real-world datasets
to verify the effectiveness of TrendNets.
For dataset construction, we chose the following international conferences: CVPR, INFOCOM, and ICASSP.
These conferences are considered prominent in the fields of computer vision, networking, and signal processing, respectively.
For each conference, we collected titles presented over the past 16 years from DBLP\footnote{https://dblp.uni-trier.de/}.
DBLP is an open bibliographic database in the fields of computer science and engineering, which provides titles, author names, publication dates, and conference/journal names, but not abstracts, for papers.
Then, from the set of words extracted from each title, we removed stop words (e.g., ``a'', ``the'', ``from'') and unnecessary symbols (e.g., ``:'', ``?'', ``!'').
Finally, we performed word stemming~\citep{80-porter-prog}.
Table~\ref{table:dataset} summarizes the details of the three datasets.

\begin{table}[t]
 \centering
\caption{
Details of each conference's dataset.
\label{table:dataset}
}
 \begin{tabular}[t]{c|c|c}\hline
  {Conference} & {\#Papers} & {\#Unique words}\\\hline
  {CVPR} & {\;7957} & {\;5468}\\
  {INFOCOM} & {\;5305} & {\;5199}\\
  {ICASSP} & {22431} & {12534}\\
  \hline
 \end{tabular}
\end{table}
We divided each conference's dataset into $T = 8$ temporal subsets
so that each time period consists of two years.
Our method was computationally efficient:
for example, matrix decomposition with $\lambda=4.0\times 10^{-4}$ on the ICASSP dataset took approximately 2.31 seconds
running on a workstation with a 3.1 GHz Intel Xeon E5-1680 Processor.

\begin{figure}[t]
\captionsetup[subfigure]{width=.94\textwidth}
\begin{minipage}{.5\textwidth}
\centering
\subfloat[TrendNets for CVPR 2011--2012.]{\label{fig:cvpr_a}\includegraphics[width=\textwidth]{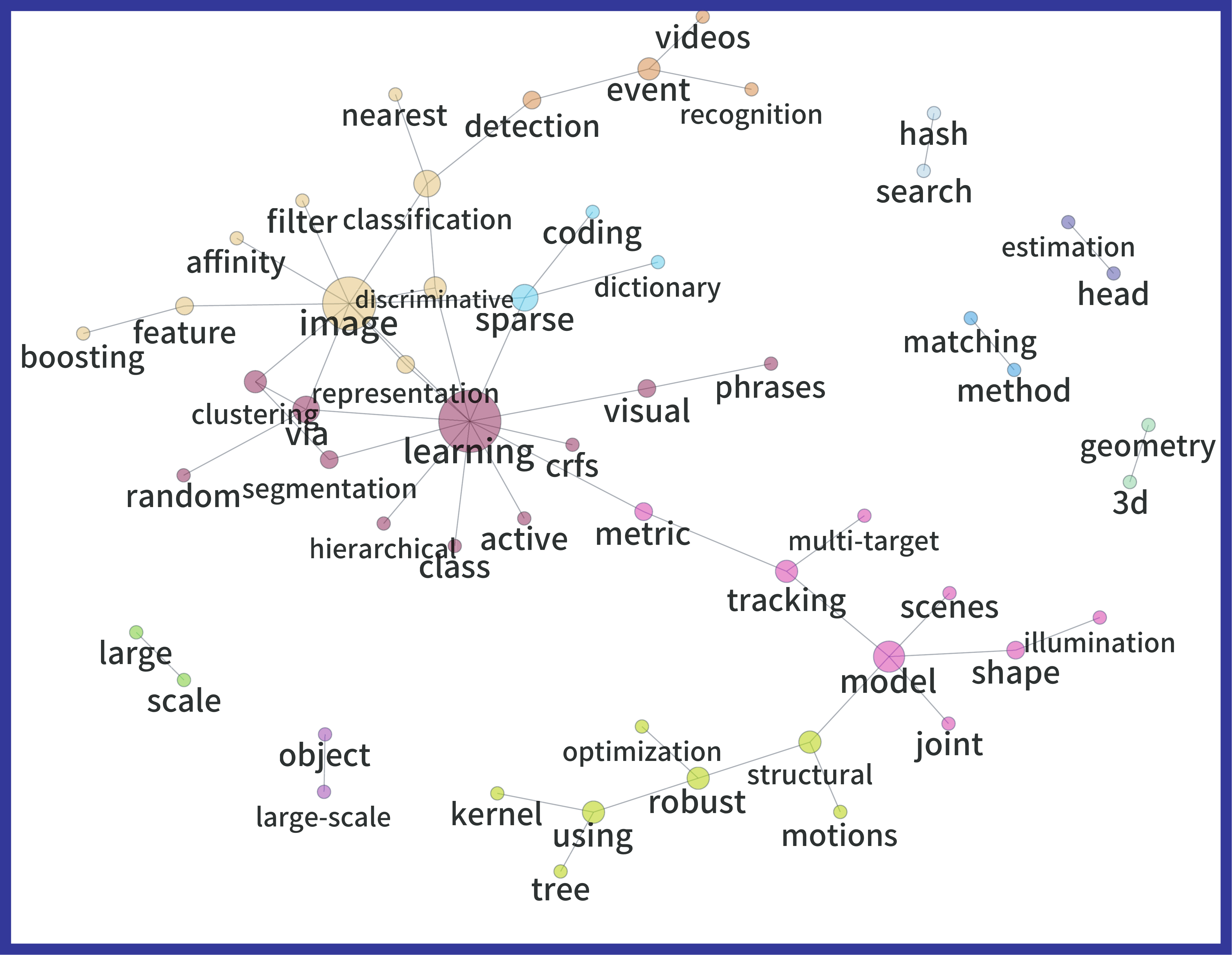}}
\end{minipage}%
\begin{minipage}{0.00\textwidth}
 \hspace{0.02\textwidth}
\end{minipage}
\begin{minipage}{.5\textwidth}
\centering
\subfloat[TrendNets for CVPR 2013--2014.]{\label{fig:cvpr_b}\includegraphics[width=\textwidth]{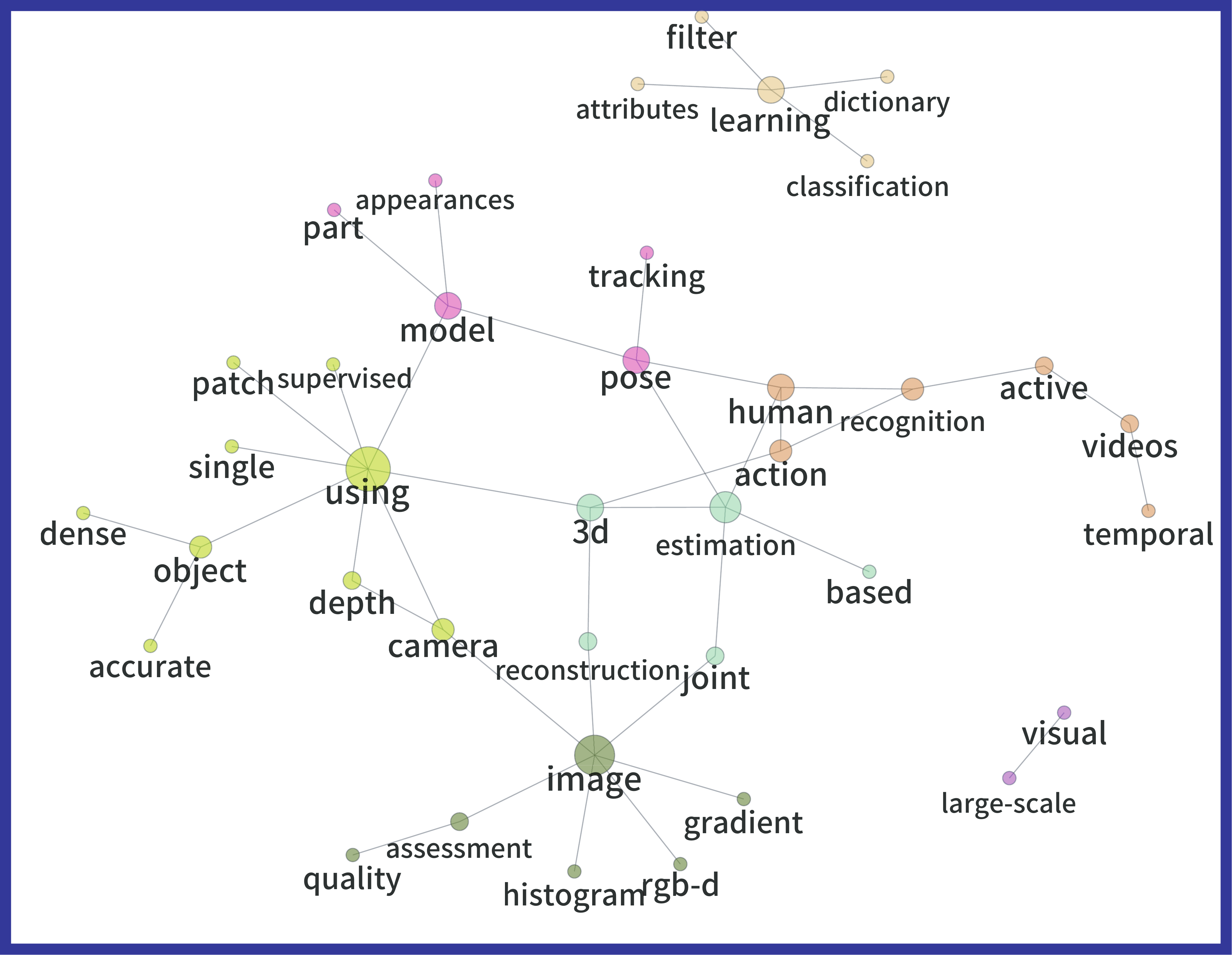}}
\end{minipage}\par\medskip
\begin{minipage}{.5\textwidth}
\centering
\subfloat[TrendNets for CVPR 2015--2016.]{\label{fig:cvpr_c}\includegraphics[width=\textwidth]{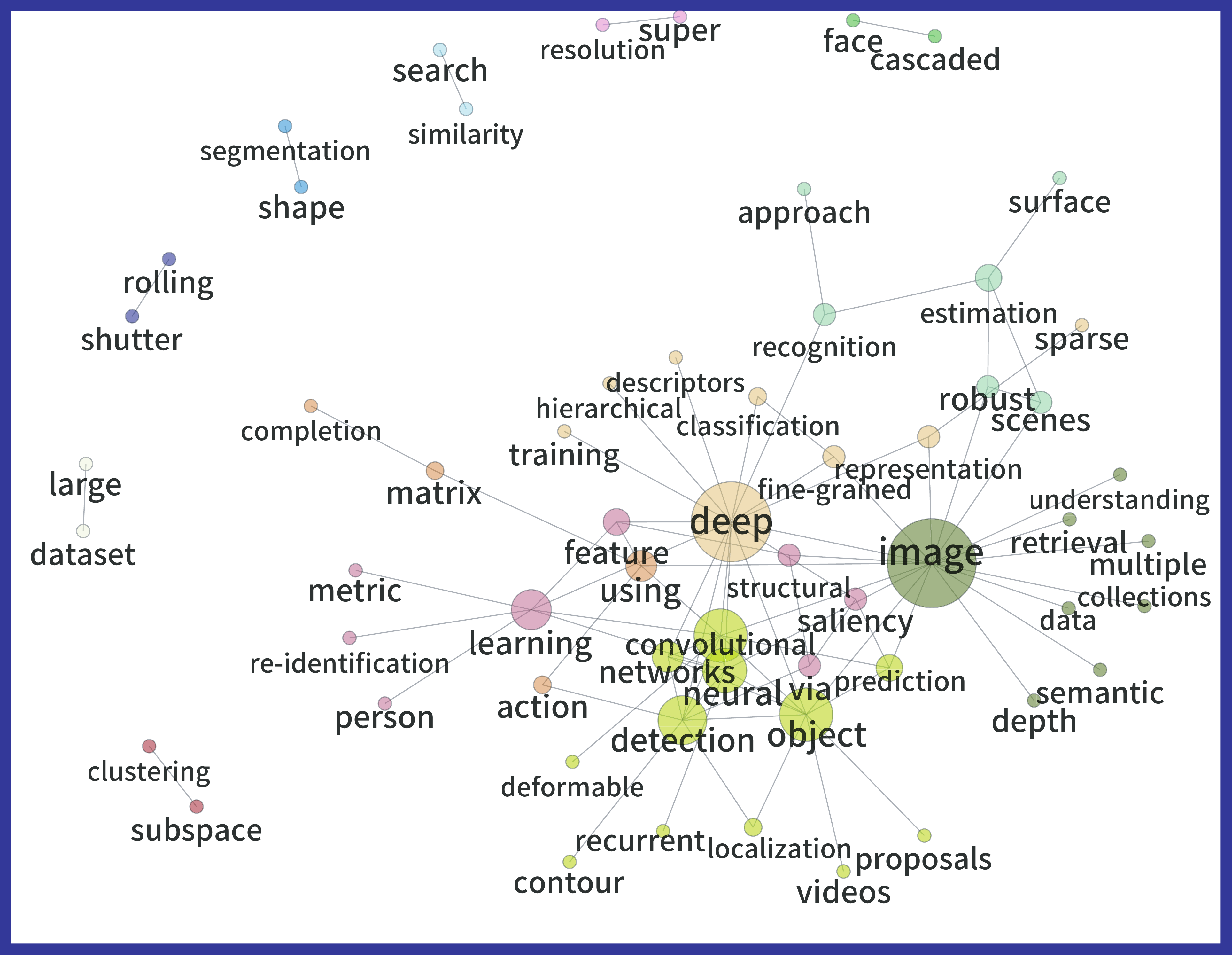}}
\end{minipage}%
\begin{minipage}{0.00\textwidth}
 \hspace{0.02\textwidth}
\end{minipage}
\begin{minipage}{.5\textwidth}
\centering
\subfloat[TrendNets for CVPR 2017--2018.]{\label{fig:cvpr_d}\includegraphics[width=\textwidth]{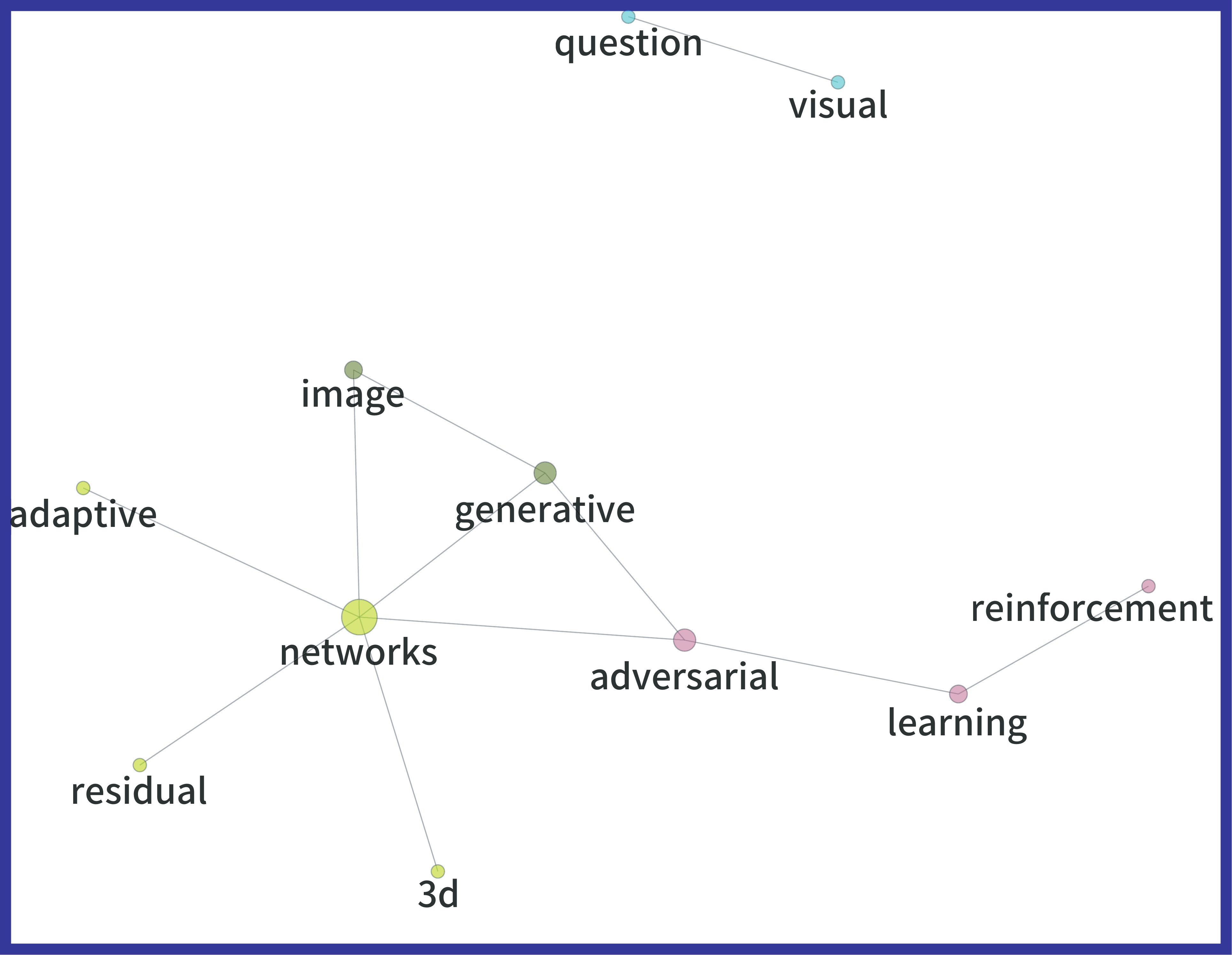}}
\end{minipage}
 \caption{
 TrendNets constructed using the CVPR dataset ($\lambda=8.5\times 10^{-6}$). (a) CVPR 2011--2012, (b) CVPR 2013--2014, (c) CVPR 2015--2016, (d) CVPR 2017--2018.
 \label{fig:cvpr}
 }
\end{figure}

Figure~\ref{fig:cvpr} shows TrendNets constructed for the latest four periods of the CVPR dataset, which used $\lambda=8.5\times 10^{-6}$, considering the limited space of the paper.
In the figure, nodes of the same color belong to the same cluster, as found via Louvain method.
To make it easier to understand research topics, each usage of words in the stemmed forms was replaced with a non-stemmed word whose co-occurrences with other words on the map were the highest. 
As shown in Fig.~\ref{fig:cvpr},
TrendNets effectively captured the characteristics of each time period in CVPR.
During 2013--2014, the appearance of ``human action recognition'' and a burst of ``pose tracking'' are captured (Fig.~\ref{fig:cvpr}b).
Figure~\ref{fig:cvpr}c contains a meaningful cluster consisting of ``convolutional neural networks'' and ``object detection,'' 
which characterizes the period from 2015 to 2016.
We can see from Fig.~\ref{fig:cvpr}d that recent emerging trends in learning are ``adversarial learning'' and ``reinforcement learning''.
A new emerging term, ``visual question,'' is also shown as an emerging trend of CVPR 2017--2018.
The results suggest that using TrendNets helps us understand topic evolution in the target conference.

\begin{figure}[t]
\begin{minipage}{.5\textwidth}
\centering
\subfloat[TrendNets for INFOCOM 2005--2006.]{\label{fig:infocom_a}\includegraphics[width=\textwidth]{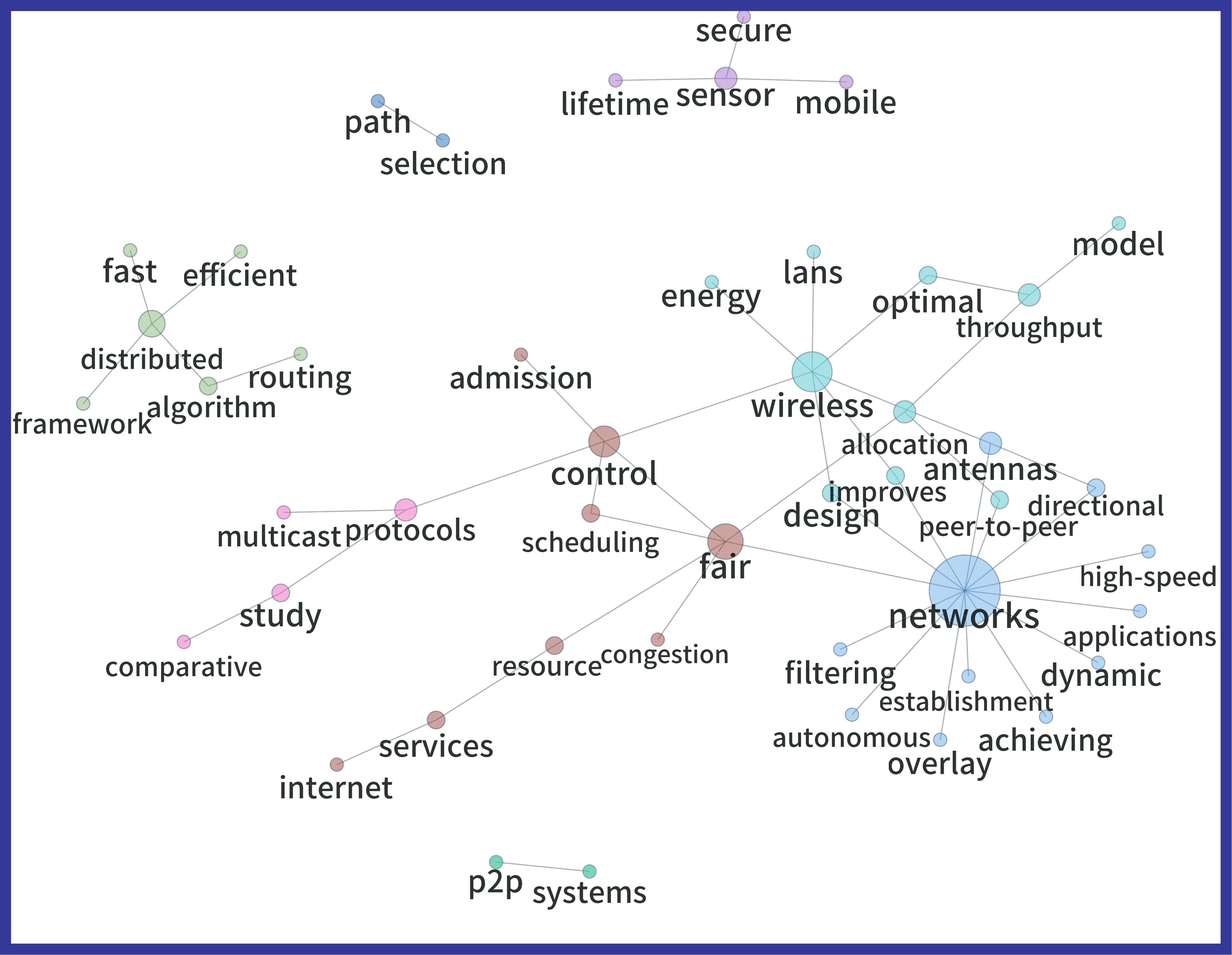}}
\end{minipage}%
\begin{minipage}{0.0\textwidth}
 \hspace{0.02\textwidth}
\end{minipage}
\begin{minipage}{.5\textwidth}
\centering
\subfloat[Original co-word network of (a).]{\label{fig:infocom_b}\includegraphics[width=\textwidth]{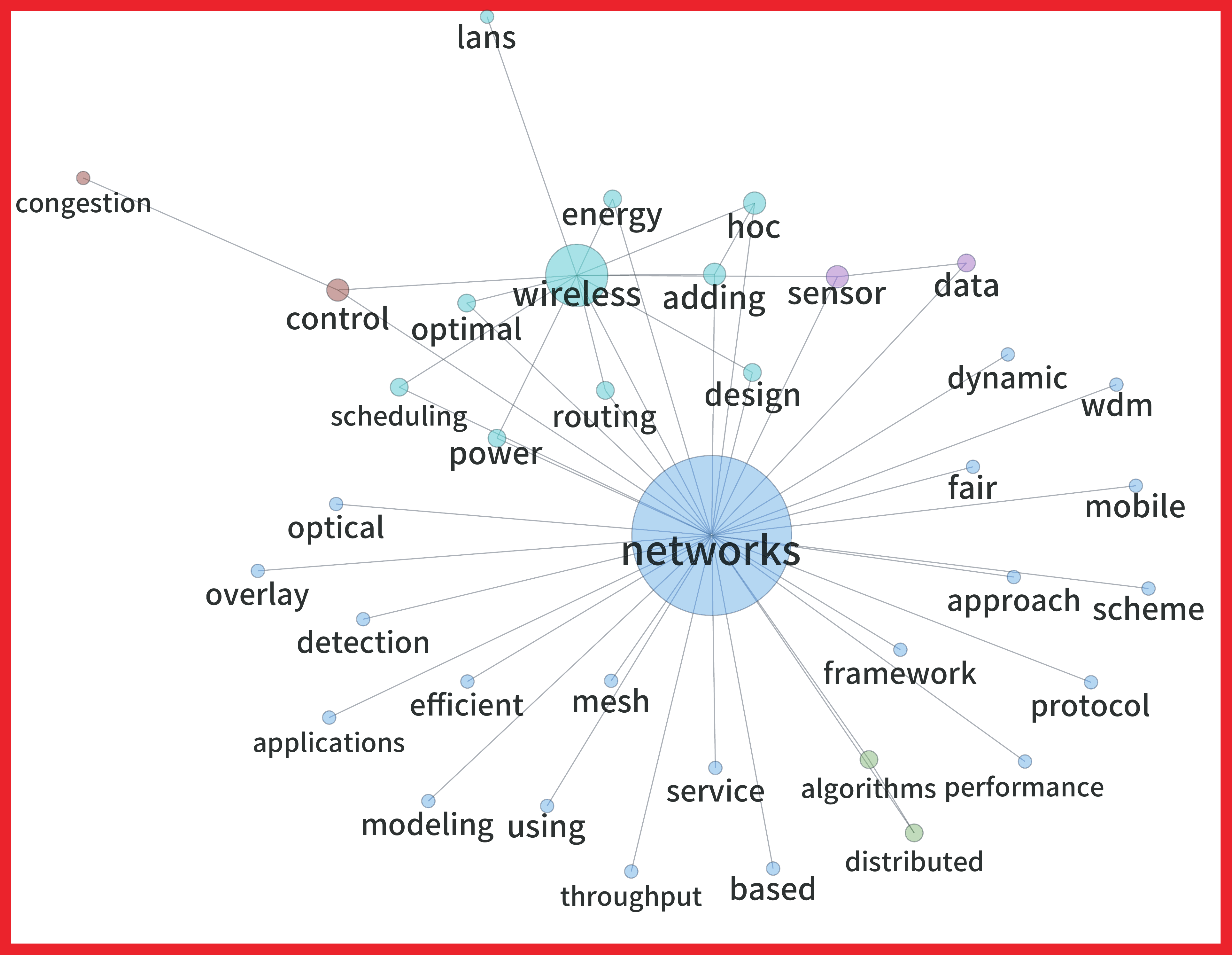}}
\end{minipage}\par\medskip
\begin{minipage}{.5\textwidth}
\centering
\subfloat[TrendNets for INFOCOM 2017--2018.]{\label{fig:infocom_c}\includegraphics[width=\textwidth]{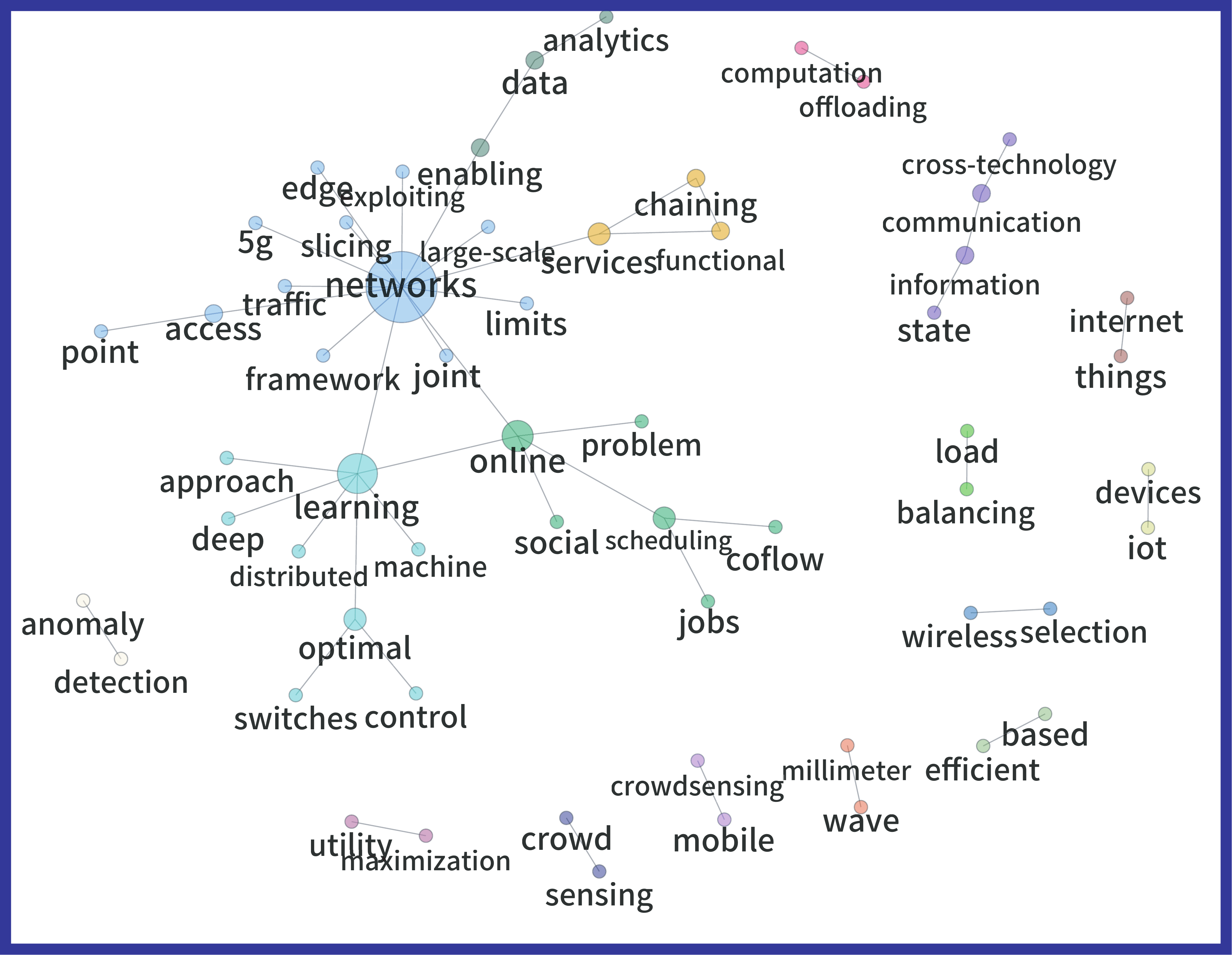}}
\end{minipage}%
\begin{minipage}{0.0\textwidth}
 \hspace{0.02\textwidth}
\end{minipage}
\begin{minipage}{.5\textwidth}
\centering
\subfloat[Original co-word network of (c).]{\label{fig:infocom_d}\includegraphics[width=\textwidth]{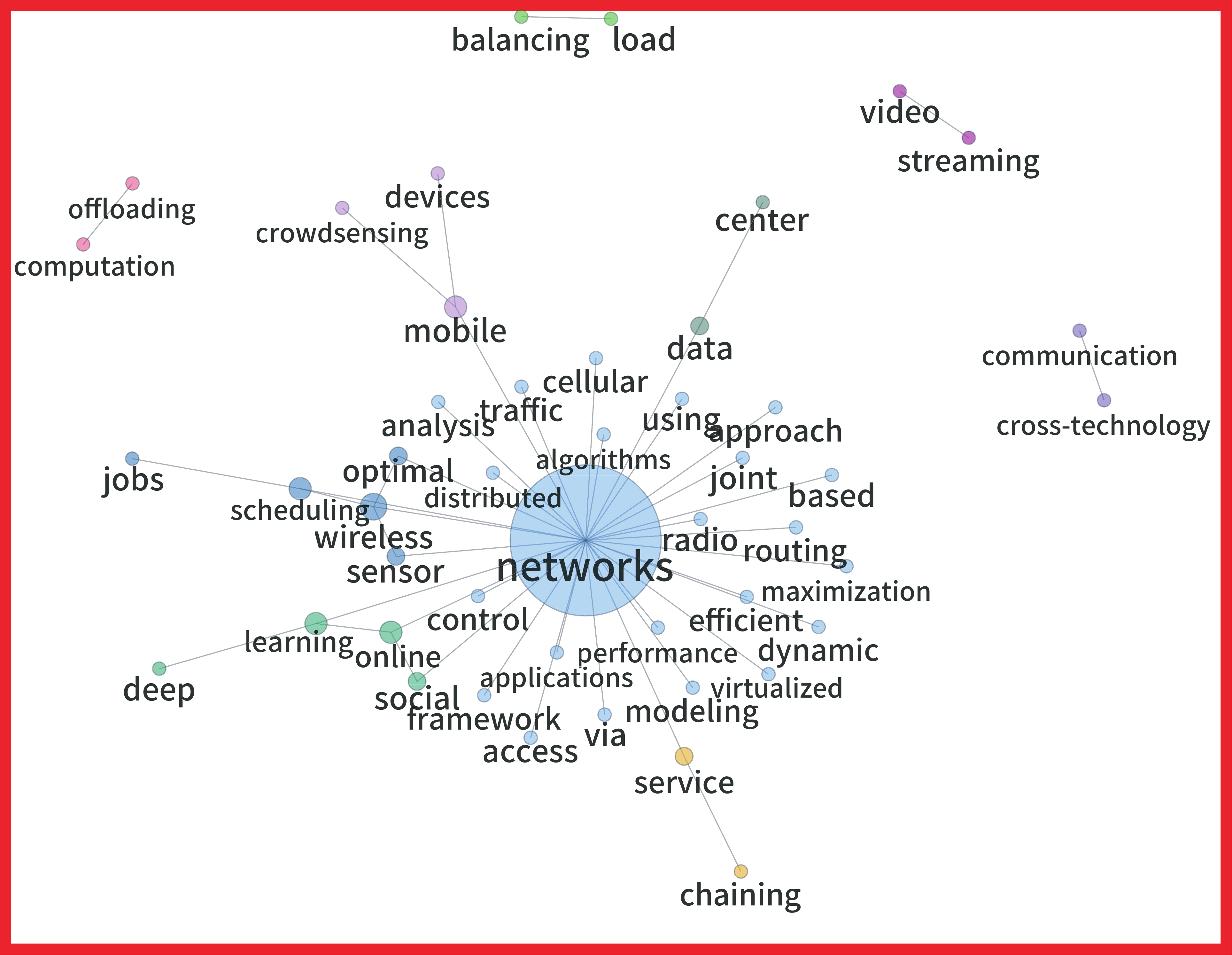}}
\end{minipage}
 \caption{
 TrendNets ($\lambda=1.7\times 10^{-3}$) and their original co-word networks for the INFOCOM dataset in a comparison of the proposed method and the traditional method. (a) TrendNets for 2005--2006, (b) original co-word network for 2005--2006, (c) TrendNets for 2017--2018, (d) original co-word network for 2017--2018. The numbers of edges of (b) and (d) were set to the same as (a) and (c) via the thresholding of edge weights, respectively.
 \label{fig:infocom}
 }
\end{figure}

\subsection*{Comparison with traditional co-word representation}

Finally, we show how TrendNets can be different from the original co-word networks
using INFOCOM and ICASSP as examples.
We performed the proposed method on the INFOCOM dataset using $\lambda=1.7\times 10^{-3}$.
Figure~\ref{fig:infocom} shows the two dynamic networks corresponding to TrendNets and the original co-word networks for INFOCOM during 2006--2007 and 2017--2018.
For fair comparison, the number of edges in the original networks were set to the same of those of the corresponding TrendNets via thresholding.
Compared with the original co-word networks in Figs.~\ref{fig:infocom}b and d,
TrendNets found multiple semantic clusters to explain the details of the bursty research topics, as shown in Figs.~\ref{fig:infocom}a and c.
Specifically, the conventional visualization (Figs.~\ref{fig:infocom}b and d) linked a lot of words to the central term ``networks,''
which makes it hard to capture the research topics.
Our method solved this conventional co-word visualization's problem and made it possible to show popular clusters 
such as ``scheduling control'' and ``learning,'' as shown in Figs.~\ref{fig:infocom}a and c, respectively.

Interestingly, as shown in Figs. 5c, 5d, 6a, and 6b, the word ``networks'' commonly appears in the two conferences’ TrendNets. Thanks to the network form-based visualization, we can infer the meaning of each conference’s ``networks'' on the basis of surrounding words. In our future work, we will detect and analyze such a difference in word usage across conferences in other fields.

\begin{figure}[t]
\begin{minipage}{.5\textwidth}
\centering
\subfloat[TrendNets for ICASSP 2015--2016.]{\label{fig:icassp_a}\includegraphics[width=\textwidth]{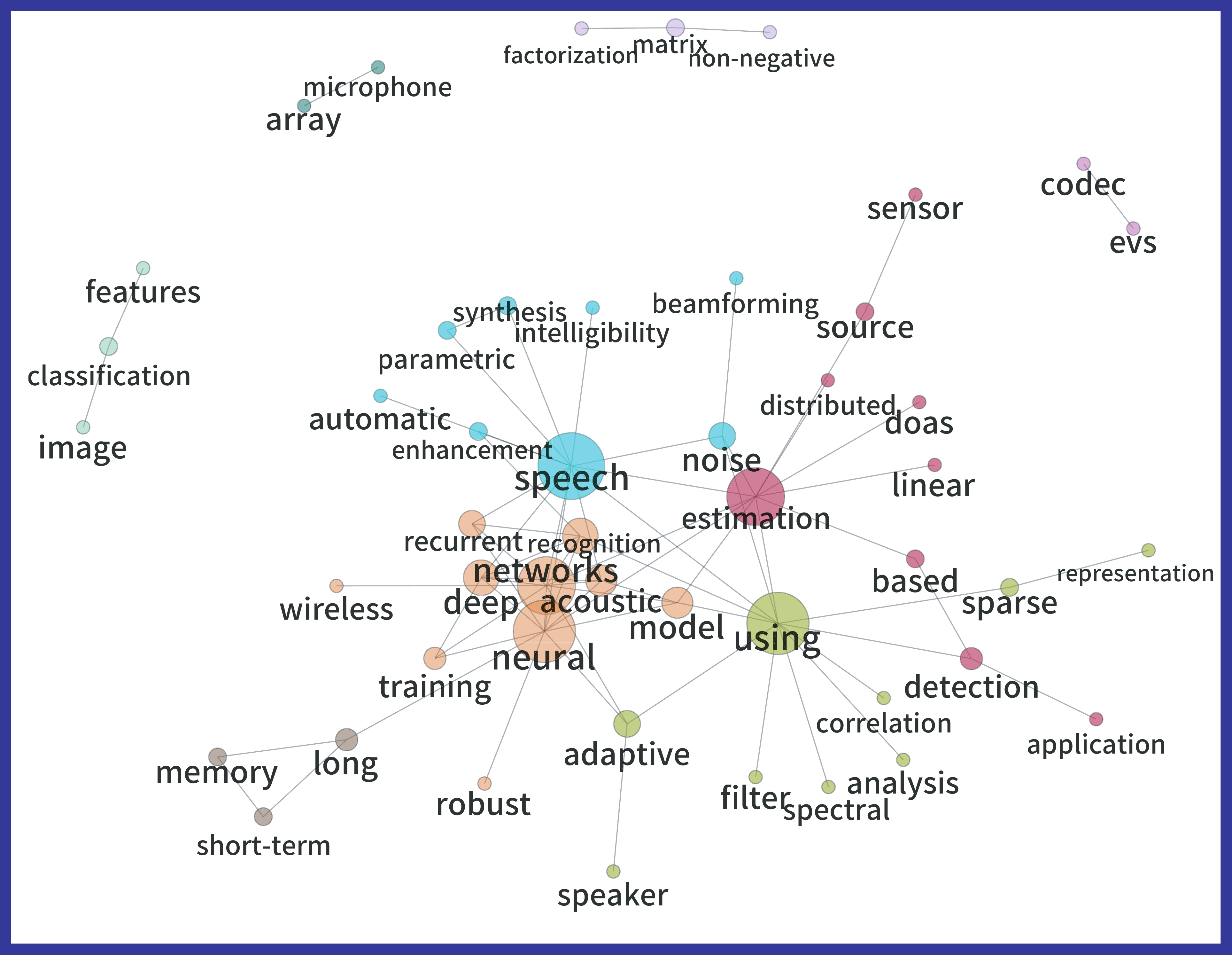}}
\end{minipage}%
\begin{minipage}{0.0\textwidth}
 \hspace{0.02\textwidth}
\end{minipage}
\begin{minipage}{.5\textwidth}
\centering
\subfloat[Original co-word network of (a).]{\label{fig:icassp_b}\includegraphics[width=\textwidth]{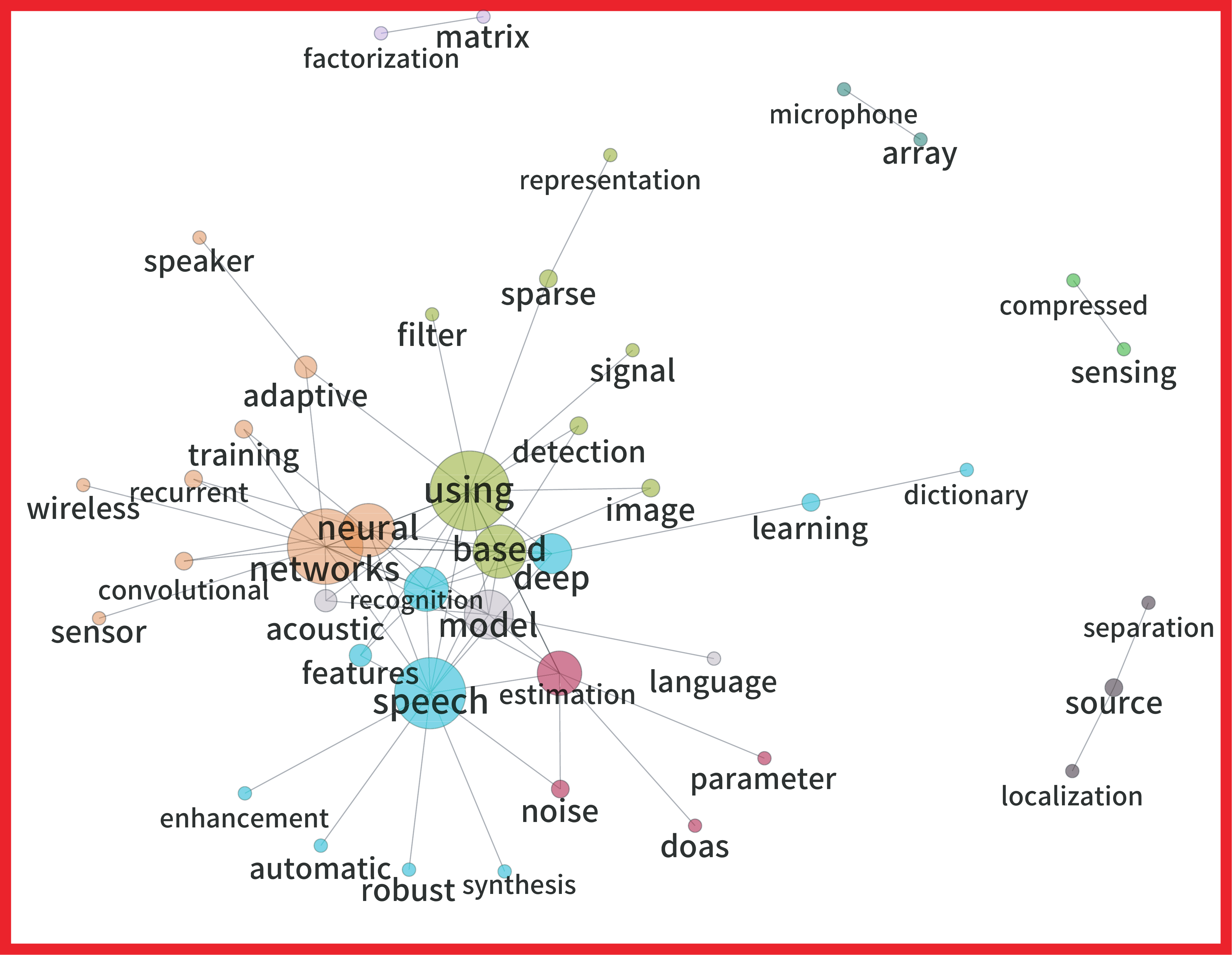}}
\end{minipage}\par\medskip
\begin{minipage}{.5\textwidth}
\centering
\subfloat[TrendNets for ICASSP 2017--2018.]{\label{fig:icassp_c}\includegraphics[width=\textwidth]{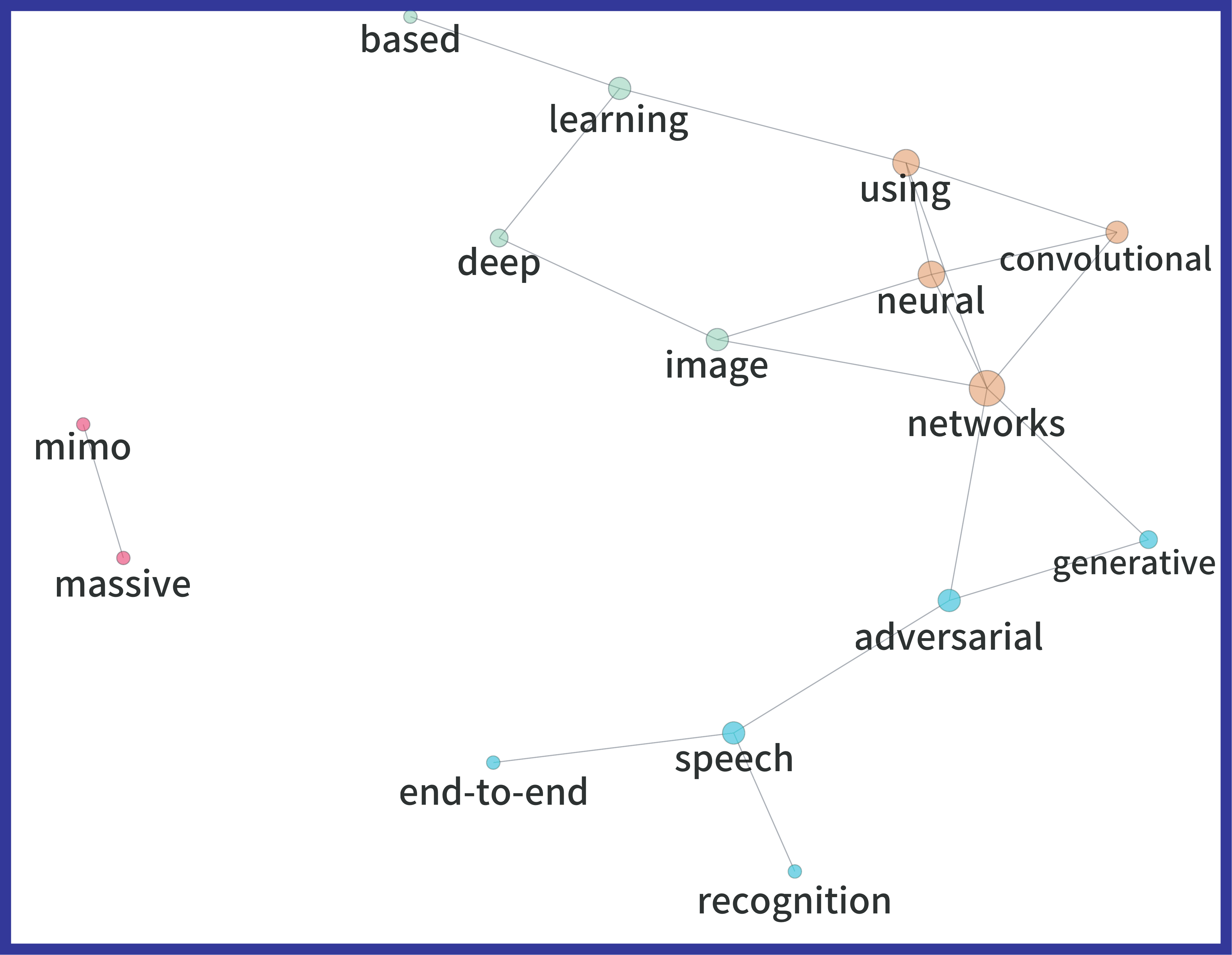}}
\end{minipage}%
\begin{minipage}{0.0\textwidth}
 \hspace{0.02\textwidth}
\end{minipage}
\begin{minipage}{.5\textwidth}
\centering
\subfloat[Original co-word network of (c).]{\label{fig:icassp_d}\includegraphics[width=\textwidth]{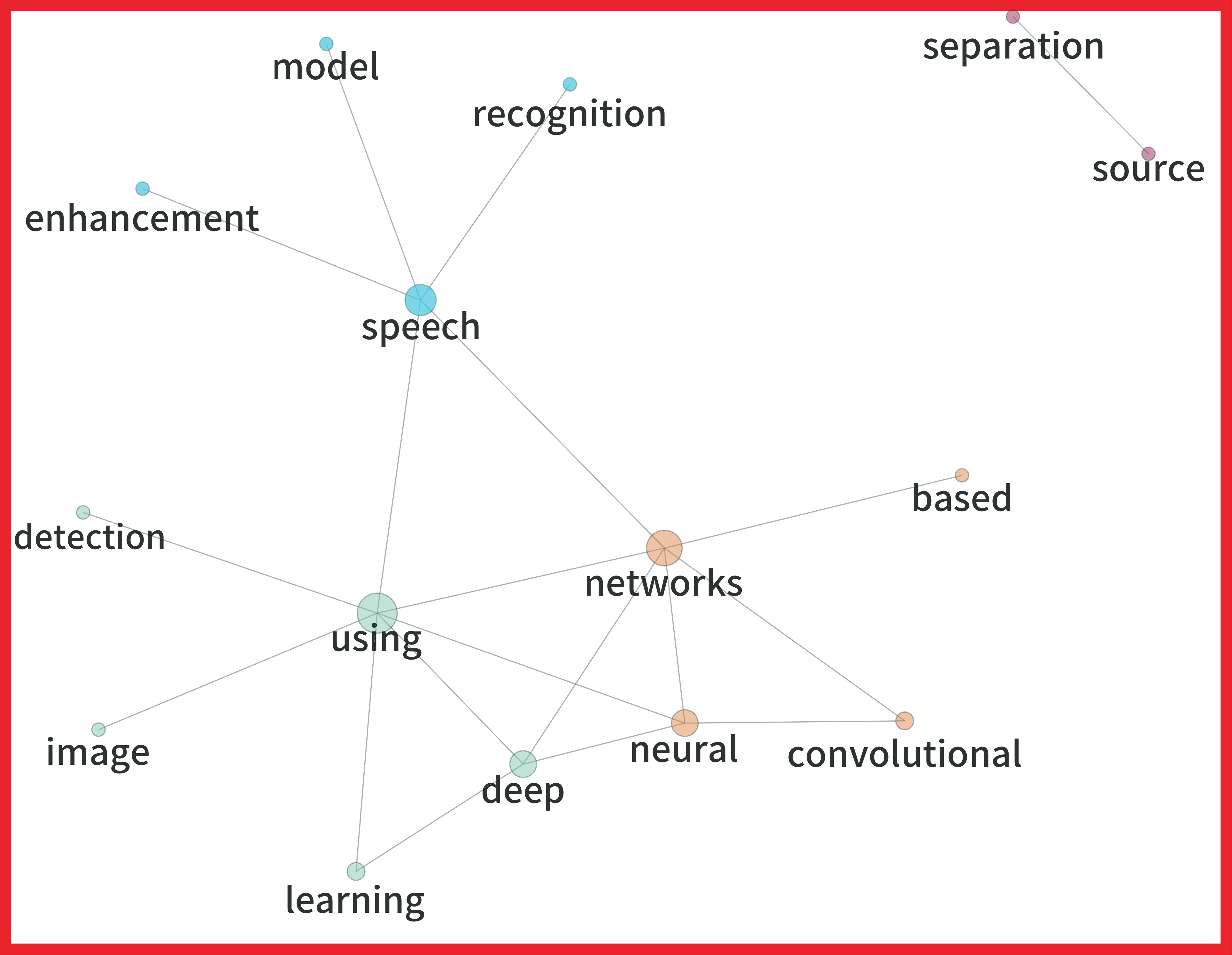}}
\end{minipage}
 \caption{
 TrendNets ($\lambda=4.0\times 10^{-4}$) and the original co-word networks for the ICASSP dataset in a comparison of the proposed method and the traditional method. (a) TrendNets for 2005--2006, (b) original co-word network for 2005--2006, (c) TrendNets for 2017--2018, (d) original co-word network for 2017--2018.
The numbers of edges of (b) and (d) were set to the same as (a) and (c) via the thresholding of edge weights, respectively.
 \label{fig:icassp}
 }
\end{figure}

We used $\lambda=4.0\times 10^{-4}$ for mapping the emerging research trends of ICASSP.
Figure~\ref{fig:icassp} shows the results on the ICASSP dataset during 2015--2016 and 2017--2018.
A large cluster of ``neural networks'' is connected to clusters of ``speech'' and ``estimation'' in Fig.~\ref{fig:icassp}a,
while the terms ``image,'' ``adversarial,'' and ``generative'' appear near neural networks in Fig.~\ref{fig:icassp}c.
In this way, TrendNets tell us that recent emerging trends of ICASSP are similar to those of CVPR,
implying that our method has the potential to find the relevance between conferences.

As compared to the INFOCOM results,
the original co-word networks seem to successfully visualize emerging trend topics.
Unfortunately, a static term ``source separation'' exists in both Figs.\ref{fig:icassp}b and d,
and more temporally representative terms can be seen in TrendNets.
Since these results of the ICASSP dataset show only a slight difference, 
we can assume that there are differences between conferences in how research topics evolve in terms of time, and it would be valuable to investigate the possibility of the proposed method to estimate it.

In summary, the proposed method enables bursty research trend visualization using a single parameter $\lambda$ only, even if they are inconspicuous in the original co-word networks.
To encourage science mapping in all research fields,
the codes for the proposed method are available on the Web\footnote{http://mm.doshisha.ac.jp/sci2/TrendNets.html}.
In addition, our TrendNets effectively revealed emerging research topics that are composed of multiple words, even though we only used the titles of papers without manually assigned keywords. Note that using abstracts in addition to the titles has the potential to cover additional words relevant to the papers that are absent from the titles.

\section*{Conclusion and Future Work}

We have developed a novel emerging research trend mapping approach that detects rapid changes in the edge weights of dynamic co-word networks.
We have formulated a convex optimization problem and provided an efficient solution to provide sparse networks named TrendNets.
Our algorithm works in time linear with the number of words in original networks.
In addition, the proposed method can easily work with a single parameter $\lambda$, which is practical and user friendly.
According to experiments using a synthetic dataset, the proposed method achieved better burst detection performance compared to other baseline methods.
Experiments conducted on three conference datasets (CVPR, INFOCOM, and ICASSP) showed the advantage of TrendNets
in detecting hot topics that are inconspicuous in the traditional co-word representation.
Interestingly, the degree of effectiveness of TrendNets differs among the conferences: we can assume that the temporal stability of a word distribution depends on the research area. Quantifying this type of characteristic is a part of our future work.

Further room for investigation and improvement exists in our work.
We currently considered only the rapid changes of the edge weights, ignoring the node degrees of original dynamic networks.
To fully make use of the original network structure for research trend visualization,
more sophisticated models need to be developed.
It is also necessary to design more effective visualization of TrendNets.
For example, we should examine how to set the colors and sizes of network components such as nodes, edges, and labels.
We will continually improve the proposed method and conduct additional experiments using a larger dataset consisting of multiple conferences/disciplines.

\begin{acknowledgements}
This research was partly supported by JSPS KAKENHI Grant Number 17K12794.
\end{acknowledgements}

\bibliographystyle{myspbasic}      
\bibliography{refs,strings,pubs}

\begin{thebibliography}{32}
\providecommand{\natexlab}[1]{#1}
\providecommand{\url}[1]{{#1}}
\providecommand{\urlprefix}{URL }
\expandafter\ifx\csname urlstyle\endcsname\relax
  \providecommand{\doi}[1]{DOI~\discretionary{}{}{}#1}\else
  \providecommand{\doi}{DOI~\discretionary{}{}{}\begingroup
  \urlstyle{rm}\Url}\fi
\providecommand{\eprint}[2][]{\url{#2}}

\bibitem[{Assefa and Rorissa(2013)}]{12-assefa-jasist}
Assefa S.~G., Rorissa A. (2013) A bibliometric mapping of the structure of
  {STEM} education using co-word analysis. Journal of the American Society for
  Information Science and Technology 64(12):2513--2536.

\bibitem[{Beck and Teboulle(2009)}]{09-beck-siam}
Beck A., Teboulle M. (2009) A fast iterative shrinkage-thresholding algorithm
  for linear inverse problems. SIAM Journal on Imaging Sciences 2:183--202.

\bibitem[{Blei and Lafferty(2006)}]{06-blei-icml}
Blei D.~M., Lafferty J.~D. (2006) Dynamic topic models. In: Proc. Int. Conf.
  Machine learning (ICML), pp. 113--120.

\bibitem[{Blei et~al.(2003)Blei, Ng, and Jordan}]{03-blei-jmlr}
Blei D.~M., Ng A.~Y., Jordan M.~I. (2003) Latent dirichlet allocation. The
  Journal of Machine Learning Research 3:993--1022.

\bibitem[{Blondel et~al.(2008)Blondel, Guillaume, Lambiotte, and
  Lefebvre}]{08-blondel-jsm}
Blondel V.~D., Guillaume J.-L., Lambiotte R., Lefebvre E. (2008) Fast unfolding
  of communities in large networks. Journal of Statistical Mechanics: Theory
  and Experiment 2008(10):P10008.

\bibitem[{B\"orner et~al.(2005)B\"orner, Dall'Asta, Ke, and
  Vespignani}]{05-borner-comp}
B\"orner K., Dall'Asta L., Ke W., Vespignani A. (2005) Studying the emerging
  global brain: {A}nalyzing and visualizing the impact of co-authorship teams.
  Complexity 10(4):57--67.

\bibitem[{Callon et~al.(1983)Callon, Courtial, Turner, and
  Bauin}]{83-callon-ssi}
Callon M., Courtial J.-P., Turner W.~A., Bauin S. (1983) From translations to
  problematic networks: An introduction to co-word analysis. Social Science
  Information 22(2):191--235.

\bibitem[{Chen(2006)}]{06-chen-jasist}
Chen C. (2006) Citespace {II}: {D}etecting and visualizing emerging trends and
  transient patterns in scientific literature. Journal of the American Society
  for Information Science and Technology 57(3):359--377.

\bibitem[{Doerfel and Barnett(1999)}]{99-doerfel-hcr}
Doerfel M.~L., Barnett G.~A. (1999) A semantic network analysis of the
  international communication association. Human Communication Research
  25(4):589--603.

\bibitem[{Drieger(2013)}]{13-drieger-pro}
Drieger P. (2013) Semantic network analysis as a method for visual text
  analytics. Procedia - Social and Behavioral Sciences 79:4--17.

\bibitem[{Hall et~al.(2008)Hall, Jurafsky, and Manning}]{08-hall-emnlp}
Hall D., Jurafsky D., Manning C.~D. (2008) Studying the history of ideas using
  topic models. In: Proc. Conf. Empirical Methods in Natural Language
  Processing (EMNLP), pp. 363--371.

\bibitem[{Hu et~al.(2013)Hu, Hu, Deng, and Liu}]{13-hu-sci}
Hu C.-P., Hu J.-M., Deng S.-L., Liu Y. (2013) A co-word analysis of library and
  information science in {C}hina. Scientometrics 97(2):369--382.

\bibitem[{Katsurai(2017)}]{17-katsurai-icbda}
Katsurai M. (2017) Bursty research topic detection from scholarly data using
  dynamic co-word networks: A preliminary investigation,. In: Proc. IEEE Int.
  Conf. Big Data Analysis (ICBDA), pp. 115--119.

\bibitem[{Kleinberg(2003)}]{03-kleinberg-kdd}
Kleinberg J. (2003) Bursty and hierarchical structure in streams. Data Mining
  and Knowledge Discovery 7(4):373--397.

\bibitem[{Li et~al.(2010)Li, Li, Cheng, Chen, Ke, Zeng, and
  Scherer}]{10-li-tits}
Li L., Li X., Cheng C., Chen C., Ke G., Zeng D.~D., Scherer W.~T. (2010)
  Research collaboration and {ITS} topic evolution: 10 years at {T-ITS}. IEEE
  Trans Intelligent Transportation Systems 11(3):517--523.

\bibitem[{Liu et~al.(2014)Liu, Goncalves, Ferreira, Xiao, Hosio, and
  Kostakos}]{14-liu-chi}
Liu Y., Goncalves J., Ferreira D., Xiao B., Hosio S., Kostakos V. (2014) {CHI}
  1994-2013: {M}apping two decades of intellectual progress through co-word
  analysis. In: Proc. ACM Conf. Human Factors in Computing Systems (CHI), pp.
  3553--3562.

\bibitem[{Mei and Zhai(2005)}]{05-mei-kdd}
Mei Q., Zhai C. (2005) Discovering evolutionary theme patterns from text: {A}n
  exploration of temporal text mining. In: Proc. ACM SIGKDD Int. Conf.
  Knowledge Discovery in Data Mining (KDD), pp. 198--207.

\bibitem[{Moreau(1962)}]{61-moreau-cr}
Moreau J.~J. (1962) Fonctions convexes duales et points proximaux dans un
  espace hilbertien. Comptes Rendus de l’Acad\'emie des Sciences de Paris
  255:2897--2899.

\bibitem[{Motter et~al.(2002)Motter, de~Moura, Lai, and
  Dasgupta}]{02-motter-phi}
Motter A.~E., de~Moura A. P.~S., Lai Y.-C., Dasgupta P. (2002) Topology of the
  conceptual network of language. Physical Review E 65:065102.

\bibitem[{Mu{\~{n}}oz-Leiva et~al.(2012)Mu{\~{n}}oz-Leiva, Viedma-del
  Jes{\'u}s, S{\'a}nchez-Fern{\'a}ndez, and
  L{\'o}pez-Herrera}]{12-munoz-quality}
Mu{\~{n}}oz-Leiva F., Viedma-del Jes{\'u}s M.~I., S{\'a}nchez-Fern{\'a}ndez J.,
  L{\'o}pez-Herrera A.~G. (2012) An application of co-word analysis and
  bibliometric maps for detecting the most highlighting themes in the consumer
  behaviour research from a longitudinal perspective. Quality {\&} Quantity
  46(4):1077--1095.

\bibitem[{Porter(1980)}]{80-porter-prog}
Porter M.~F. (1980) An algorithm for suffix stripping. Program 3(14):130--137.

\bibitem[{Ravikumar et~al.(2015)Ravikumar, Agrahari, and
  Singh}]{15-ravikumar-sci}
Ravikumar S., Agrahari A., Singh S.~N. (2015) Mapping the intellectual
  structure of scientometrics: a co-word analysis of the journal scientometrics
  (2005--2010). Scientometrics 102(1):929--955.

\bibitem[{Ronda-Pupo and Guerras-Martin(2012)}]{12-ronda-smj}
Ronda-Pupo G.~A., Guerras-Martin L.~A. (2012) Dynamics of the evolution of the
  strategy concept 1962–2008: a co-word analysis. Strategic Management
  Journal 33(2):162--188.

\bibitem[{Shibata et~al.(2008)Shibata, Kajikawa, Takeda, and
  Matsushima}]{08-shibata-tech}
Shibata N., Kajikawa Y., Takeda Y., Matsushima K. (2008) Detecting emerging
  research fronts based on topological measures in citation networks of
  scientific publications. Technovation 28(11):758--775.

\bibitem[{Silva et~al.(2016)Silva, Amancio, Bardosova, Costa, and
  Jr.}]{16-silva-ji}
Silva F.~N., Amancio D.~R., Bardosova M., Costa L.~F., Jr. O. N.~O. (2016)
  Using network science and text analytics to produce surveys in a scientific
  topic. Journal of Informetrics 10(2):487--502.

\bibitem[{Song et~al.(2014)Song, Heo, and Kim}]{14-song-sci}
Song M., Heo G.~E., Kim S.~Y. (2014) Analyzing topic evolution in
  bioinformatics: investigation of dynamics of the field with conference data
  in {DBLP}. Scientometrics 101(1):397--428.

\bibitem[{Topalli and Ivanaj(2016)}]{16-topalli-jwb}
Topalli M., Ivanaj S. (2016) Mapping the evolution of the impact of economic
  transition on {C}entral and {E}astern {E}uropean enterprises: {A} co-word
  analysis. Journal of World Business
  \doi{http://dx.doi.org/10.1016/j.jwb.2016.06.003}.

\bibitem[{Vlachos et~al.(2004)Vlachos, Meek, Vagena, and
  Gunopulos}]{04-vlachos-sigmod}
Vlachos M., Meek C., Vagena Z., Gunopulos D. (2004) Identifying similarities,
  periodicities and bursts for online search queries. In: Proc. ACM SIGMOD Int.
  Conf. Management of Data (SIGMOD), pp. 131--142.

\bibitem[{Wang et~al.(2008)Wang, Blei, and Heckerman}]{08-wang-uai}
Wang C., Blei D.~M., Heckerman D. (2008) Continuous time dynamic topic models.
  In: Proc. Int. Conf. Uncertainty in Artificial Intelligence (UAI), pp.
  579--586.

\bibitem[{Wang and McCallum(2006)}]{06-wang-kdd}
Wang X., McCallum A. (2006) Topics over time: {A} non-markov continuous-time
  model of topical trends. In: Proc. Int. Conf. Knowledge Discovery and Data
  Mining (KDD), pp. 424--433.

\bibitem[{Wang et~al.(2014)Wang, Cheng, and Lu}]{14-wang-sci}
Wang X., Cheng Q., Lu W. (2014) Analyzing evolution of research topics with
  {NEV}iewer: a new method based on dynamic co-word networks. Scientometrics
  101(2):1253--1271.

\bibitem[{Zhang et~al.(2012)Zhang, Xie, Hou, Tu, Xu, Song, Wang, and
  Lu}]{12-zhang-plos}
Zhang J., Xie J., Hou W., Tu X., Xu J., Song F., Wang Z., Lu Z. (2012) Mapping
  the knowledge structure of research on patient adherence: Knowledge domain
  visualization based co-word analysis and social network analysis. PLoS ONE
  7:1--7.

\end{thebibliography}

\end{document}